\documentclass[12pt,singlespace]{article}
\usepackage{putex}
\pdfoutput=1
\usepackage{graphicx}

\usepackage{hyperref}
\usepackage{subfigure}
\usepackage{graphicx}
\usepackage{enumerate}
\usepackage{cite}
\usepackage{tensor}
\usepackage{slashed}
\usepackage{placeins}
\usepackage{bigints}

\numberwithin{equation}{section}

\newcommand {\be} {\begin {equation}}
\newcommand {\ee} {\end {equation}}

\newcommand {\bes} {\begin {equation*}}
\newcommand {\ees} {\end {equation*}}

\newcommand{\es}[2] {\begin{equation} \label{#1} \begin{split} #2 \end{split} \end{equation}}

\newcommand{\beq}{\begin{equation}}
\newcommand{\eeq}{\end{equation}}

\newcommand {\kms} {\,\,\text{km}/\text{s}}

\newcommand{\V}[1]{{ \bf #1}}
\newcommand{\obs}{\text{obs}}
\newcommand{\Sun}{\odot}
\newcommand{\earth}{\oplus}
\newcommand{\Min}{\text{min}}
\newcommand{\erf}{\text{erf}}
\newcommand{\esc}{\text{esc}}

\newcommand{\kevnr}{\, \, \text{keV}_\text{nr}}

\begin{document}

\institution{PCTS}{Princeton Center for Theoretical Science, Princeton University, Princeton, NJ 08544}
\institution{PU}{Department of Physics, Princeton University, Princeton, NJ 08544}

\title{
Dark-Matter Harmonics Beyond Annual Modulation
}

\authors{Samuel K. Lee\worksat{\PCTS}, Mariangela Lisanti\worksat{\PCTS,\PU}, and Benjamin R.~Safdi\worksat{\PU}
}

\abstract{
The count rate at dark-matter direct-detection experiments should modulate annually due to the motion of the Earth around the Sun.    
We show that higher-frequency modulations, including daily modulation, are also present and in some cases are nearly as strong as the annual modulation.  These higher-order modes are particularly relevant if $(i)$ the dark matter is light, $O(10)$ GeV, $(ii)$ the scattering is inelastic, or $(iii)$ velocity substructure is present; for these cases, the higher-frequency modes are potentially observable at current and ton-scale detectors.  We derive simple expressions for the harmonic modes as functions of the astrophysical and geophysical parameters describing the Earth's orbit, using an updated expression for the Earth's velocity that corrects a common error in the literature.  For an isotropic halo velocity distribution, certain ratios of the modes are approximately constant as a function of nuclear recoil energy.  Anisotropic distributions can also leave observable features in the harmonic spectrum.  Consequently, the higher-order harmonic modes are a powerful tool for identifying a potential signal from interactions with the Galactic dark-matter halo.
}

\maketitle

\section{Introduction}  \label{mag}

Perhaps the most concrete evidence for physics beyond the Standard Model is the observation that the Universe consists of approximately five times as much dark matter (DM) as baryonic matter~\cite{Komatsu:2010fb,Ade:2013zuv}.  Little is known about the dark sector beyond its gravitational interactions.  In particular, it is unknown how dark matter couples to ordinary matter non-gravitationally, or if it does so at all.  Direct-detection experiments~\cite{Goodman:1984dc, Drukier:1983gj} that seek to observe collisions of DM particles with nuclei in underground detectors~\cite{Smith:1988kw,Bertone:2004pz,Jungman:1995df,Gaitskell:2004gd,Freese:2012xd} could potentially shed light on the properties of the DM and its interactions with ordinary matter. 

The differential scattering rate of a DM particle of mass $m_\chi$ scattering off a nucleus 
with momentum transfer $q$ and an effective cross section $\sigma(q^2)$ is
\es{recoilR}{
{ d R \over d E_{\text{nr}} } = {\rho_\chi \over 2 m_\chi \mu^2} \,\, \sigma(q^2)  \,\,\eta( v_{\text{min}}, t)\ , \quad \text{with} \quad  \eta(v_{\text{min}},t ) = \bigintssss_{v_{\text{min}}}^{\infty}\frac{\tilde f\big( \V{v} + \V{v}_\text{obs}(t)\big)}{v}\,d^3v \,.
}
Here, $E_\text{nr}$ is the nuclear recoil energy, $\eta(v_{\text{min}},t)$ is the mean inverse lab-frame speed of the DM particles, $\mu$ is the reduced mass of the DM-nucleus system, $\rho_\chi$ is the local DM density, and $v_{\text{min}}$ is the threshold speed (\emph{i.e.}, the minimum lab-frame DM speed required to yield a recoil energy $E_\text{nr}$).  The DM velocity distribution in the Galactic frame is $\tilde f(\V{v})$, which is boosted to the lab frame by shifting all velocities by $\V{v}_\obs(t)$, the detector's velocity in the Galactic frame.  The scattering rate varies annually as a result of the Earth's orbit about the Sun~\cite{Drukier:1986tm}, and it also varies daily due to the Earth's rotation about its axis~\cite{Freese:2012xd}.  The annual modulation is due to the fact that the Galactic DM halo appears as a constant ``wind" in the Solar reference frame.  As the Earth travels through its orbit, it sometimes heads into the wind, leading to a larger DM flux, and sometimes away, leading to a smaller flux.  The daily modulation has the same intuitive explanation, although it is a subdominant effect.   
  \begin{figure}[tb]
  \leavevmode
\begin{center}$
\begin{array}{cc}
\scalebox{1.7}{\includegraphics{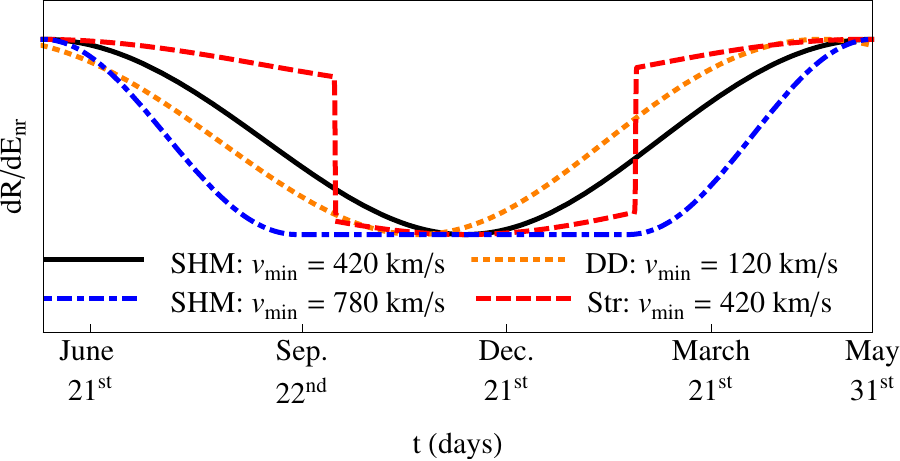}}  
  \end{array}$
\end{center}
\caption{The annual variation of the differential rate is sensitive to the threshold speed $v_\Min$ and the distribution of DM velocities.  The Standard Halo Model (SHM) distribution is taken to have $v_0 = 220 \kms$ and $v_\esc = 550 \kms$, the dark disk (DD) is a pure Maxwellian distribution with $v_0 = 70 \kms$ and $v_\text{lag} = 50 \kms$, and the stream (Str) travels at a speed of $350 \kms$ towards the north Galactic pole.  For the dark disk and stream, the densities are $\rho_\text{D} = \rho_\text{SHM}$ and $\rho_\text{Str} = 0.1 \rho_\text{SHM}$ for definiteness, where $\rho_\text{SHM} = \rho_\chi$ is the density of the SHM component.  To emphasize the differences in shape, we have shifted and scaled the differential rates so that they share the same maximum and minimum values.}
\label{AnMod}
\end{figure}

In Fig.~\ref{AnMod}, the solid-black curve shows the characteristic sinusoidal time dependence of the differential rate -- assuming a Standard Halo Model (SHM) velocity distribution~\cite{Freese:2012xd}, defined in Appendix~\ref{SHM} -- when the lab-frame threshold speed is low.  In this case, scattering is typically induced by DM particles with Galactic-frame velocities well below the Galactic-frame escape velocity, $v_\text{esc}$.  However, extremely high lab-frame threshold speeds may require Galactic-frame velocities in excess of $v_\text{esc}$; at such velocities, the distribution function is unpopulated.  Thus, at sufficiently high threshold speeds, this unpopulated region will be probed for some portion of the year as a result of the variation of the Earth's velocity, causing the differential rate to ``turn off" during this time; this is illustrated in Fig.~\ref{AnMod} by the dashed-dotted blue curve.  The strength of this turn-off effect depends on the details of the high-velocity tail of the DM velocity distribution.  Furthermore, the other curves in Fig.~\ref{AnMod} show how the time dependence can vary in the presence of velocity substructure such as a DM stream (dashed red) or disk (dotted orange).    Figure~\ref{AnMod} thus shows that varying the assumptions about the DM velocity distribution can dramatically alter the time dependence of a signal.  Non-sinusoidal modulation of the differential rate has been discussed in the context of the SHM plus DM streams~\cite{Gelmini:2000dm, Freese:2003na,Savage:2006qr} and anisotropic halos~\cite{Fornengo:2003fm,Green:2003yh}. 

To capture the deviations from a pure sinusoid, the differential scattering rate may be expanded as a Fourier series 
\es{drdEtyp}{
{ d R \over d E_{\text{nr}} } = A_0 + \sum_{n = 1}^\infty\big[ A_n \cos n\, \omega (t - t_0) +  B_n \sin n \, \omega (t - t_0) \big]\,
}
for a fixed period of one year, where $\omega = 2 \pi / \text{year}$ and $t_0$ is a phase parameter.  $A_0$ is the unmodulated rate, and $A_1$ is typically referred to as the annual modulation.  All other $A_n$ and $B_n$ are higher-order Fourier components.  This expansion is useful when looking for effects due to the Earth's orbit around the Sun.  When studying the daily modulation, the following is instead useful
\es{dailyMod}{
{ d R \over d E_{\text{nr}} } \approx A_0 + A_d \cos \big[ \omega_d\, (t - t_0^d) + \lambda_0 \big] \,,
}  
where $\omega_d = 2 \pi / \text{day}$ is the {\it sidereal} daily frequency (a sidereal day is approximately 23 hours and 56 minutes), $t_0^d$ is a phase parameter, and $\lambda_0$ is the longitude of the experiment on the surface of the Earth.  Note that this effect is different from what is observed by directional detectors~\cite{Spergel:1987kx}.  There, the~\emph{direction} of the nuclear track modulates daily due to the reversal of the detector relative to the DM halo as the Earth completes a rotation about its axis.  In contrast,~\eqref{dailyMod} is the change in the \emph{overall} scattering rate (integrated over all angles) due to the Earth's daily rotation.

The purpose of this work is to explore in detail the behavior and ramifications of the higher-order harmonic components in DM scattering, extending the first steps taken in this direction by~\cite{Alves:2010pt,Chang:2011eb}.  For the most common assumptions about the dark sector, $|A_1| / A_0 \ll 1$, and the higher-order Fourier components are even further suppressed.  However, we find that in many realistic scenarios -- including, most notably, that of $O$(10) GeV DM -- many current and next-generation experiments are sensitive to the higher-order modes.
  
Specifically, the leading modes beyond annual modulation are $A_2$, $B_1$, $B_2$ and the daily modulation, $A_d$.  These modes are related to unique features of the Earth's trajectory, including the speed and eccentricity of its orbit, the alignment of the ecliptic plane with respect to the Galactic plane, the rotational speed of the Earth, the obliquity of the Earth's axis, and the latitude and longitude of the experiment.  Under some assumptions about the DM halo, the ratios of many of the harmonic modes -- such as $B_1 / A_1$, $B_2 / A_1$, and $A_d / A_1$ -- are constant with nuclear recoil energy, and the constant values are fixed by astrophysical and geophysical parameters.  Because the properties of the Earth's motion leave a unique fingerprint on the harmonic spectrum, higher-order modes are a consistency check for a potential DM signal.
     
Additionally, the higher-order Fourier modes probe astrophysical and particle-physics effects -- \emph{e.g.}, particle-physics models sensitive to large threshold speeds.  They can help determine the DM mass and constrain scenarios such as inelastic scattering.  In addition, as alluded to in Fig.~\ref{AnMod}, these modes are affected by the presence of Galactic velocity substructure, such as the dark disk and streams.  As an example, a DM stream with a density of a few percent the density of the bulk Galactic halo has little effect on the unmodulated rate, a small effect on the annual modulation, and a profound effect on the higher-order harmonic modes.  For this reason, experiments that observe an annual modulation can immediately start constraining a variety of astrophysical and particle-physics scenarios using the higher-order harmonic analysis.     

The remainder of this paper proceeds as follows.  Section~\ref{sec: Scaling} introduces basic scaling relations to build intuition on the higher-order harmonic modes that dominate for specific astrophysical and particle-physics scenarios.  Sections~\ref{sec: mis} and~\ref{sec: dailymodulation} present the complete derivations for the Fourier modes and daily modulation, respectively.  The implications of these results are presented in Sec.~\ref{sec: examples}.  We conclude in Sec.~\ref{sec: conclusions}.  Appendix~\ref{poisson} describes the statistical procedure for determining the amount of exposure needed to see a particular harmonic mode.  A derivation of the Earth's velocity is given in Appendix~\ref{sec: ORBIT} and corrects a commonly used, erroneous result in the literature~\cite{Lewin:1995rx}.  Analytic results for various models of the DM velocity distribution are given in Appendix~\ref{SHM}.

\section{Scaling Relations for Higher-Order Harmonics} 
\label{sec: Scaling}

This section begins to explore the higher-order terms in the Fourier expansion of the differential scattering rate given in~\eqref{drdEtyp}.  In particular, we want to build intuition for the exposure needed to observe a mode $A_n$ or $B_n$ relative to another mode, for particular properties of the DM.  The exposure is the total integrated amount of detector mass times exposure time, and we use $E(A_n)$ and $E(B_n)$ to denote the exposure required to detect a given mode for specific DM model and experimental parameters.  
In a background-free experiment,  the amount of exposure needed to detect a mode with 95\% confidence is
\es{EAEBapprox}{
E(A_0) &\approx  \left[ \int_{E_\text{thresh}}^{E_\text{max}} d E_\text{nr} {A_0(E_\text{nr}) } \right]^{-1}  \,, \qquad E(A_n) \approx 7.68\left[ \int_{E_\text{thresh}}^{E_\text{max}} d E_\text{nr} {A_n^2(E_\text{nr}) \over A_0(E_\text{nr})} \right]^{-1} \,.
}
For a complete derivation, see Appendix~\ref{poisson}.  The corresponding expression for $B_n$ is similar, with the obvious substitutions.  

In many cases, the magnitude of the higher-order modes scales simply with the parameter
\es{epsilonDefG}{
\epsilon \equiv {V_\earth \over 4 v_\Sun} \approx {7.4 \kms \over v_\text{rot}} \left( 1 - {12 \kms \over v_\text{rot}} + \cdots \right)\,,
}
where $v_\Sun = | \V{v}_\Sun|$ is the speed of the DM reference frame relative to the Sun's reference frame, and $V_\earth = |\V{V}_\earth| \approx 29.79 \kms $ is the speed of the Earth's orbit.  
When the DM rest frame coincides with that of the Galaxy,
${\bf v}_{\odot} = {\bf v}_{\text{LSR}} + {\bf v}_{\odot, \text{pec}}$, with $ \V{v}_{\text{LSR}} = (0, v_{\text{rot}},0)$, $v_\text{rot} = 220 \kms$, and $\V{v}_{\odot,\text{pec}} = (11,12,7) \kms$~\cite{Schoenrich:2009bx,2000A&A...354..522M}.\footnote{In Galactic coordinates, the ${\bf \hat x}$ axis points towards the center of the Galaxy, the ${\bf \hat y}$ axis points in the direction of the local disk rotation, and the ${\bf \hat z}$ axis is normal to the Galactic plane and points toward the north Galactic pole.}   
For a velocity distribution that is smooth and isotropic, the modes  follow the scaling relations
\begin{equation} \label{AnBnEpsilon}
A_n / A_0 \sim \epsilon^n \quad \text{ and } \quad  B_n / A_0 \sim \epsilon^n,
\end{equation}
as will be derived in Sec.~\ref{sec: mis}.  For a standard isotropic distribution whose reference frame coincides with that of the Galaxy, $\epsilon \approx 0.032$ and the higher-order modes are strongly suppressed.  

Note that the scaling of the mode $B_1$ is a notable exception to~\eqref{AnBnEpsilon}.  As will be shown later, $B_1 / A_0 \sim \epsilon^{2}$ due to the eccentricity of the Earth's orbit.  In the case of the daily-modulation mode $A_d$,
\es{AdA0}{
{A_d \over A_0} \sim  {V_\text{d} \over v_\Sun} \approx 2 \times 10^{-3} \,,
}
where $V_\text{d} \approx 0.46 \kms$ is the rotational speed of the Earth at the equator.  Coincidentally, the ratio in~\eqref{AdA0} is $O(\epsilon^2)$, so the ease of observing daily modulation is roughly similar to observing the modes $A_2, B_1,$ and $B_2$.

\begin{figure}[tb]
  \leavevmode
\begin{center}$
\begin{array}{cc}
\scalebox{1.75}{\includegraphics{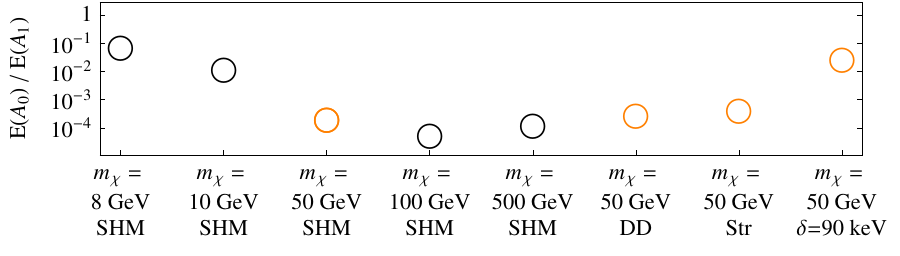}}  
  \end{array}$
\end{center}
\vspace{-0.3in}
\caption{The exposure needed to observe the unmodulated rate $A_0$ relative to that needed to observe the annual modulation $A_1$ at  XENON1T.  This ratio is enhanced relative to that for DM heavier than 50  GeV in the SHM if the DM is $(i)$ $O$(10) GeV, $(ii)$ in a disk (DD) or stream (Str), or $(iii)$ inelastic with a mass splitting of $\delta=90$ keV; cases $(i)$ and $(iii)$ are most significant.  For the dark disk and stream, the densities are $\rho_\text{D} = \rho_\text{SHM}$ and $\rho_\text{Str} = 0.1 \rho_\text{SHM}$ for definiteness. }  
\label{A0vA1}
\end{figure}

Figure~\ref{A0vA1}
 shows the projection for  $E(A_0)/E(A_1)$ at the XENON1T experiment\footnote{Throughout this paper, we consider ideal, background-free Xenon and Germanium detectors as canonical examples.  The Xenon detector is roughly modeled after XENON1T~\cite{Alfonsi:2012bfa}, with $(E_\text{thresh}, E_\text{max}) = (4, 50)$ keV$_\text{nr}$.  The Germanium detector is modeled after GEODM DUSEL~\cite{FigueroaFeliciano:2010zz}  with $(E_\text{thresh}, E_\text{max}) = (5, 100)$ keV$_\text{nr}$.  The planned fiducial masses of XENON1T and GEODM DUSEL are around 1 ton and 1.5 tons, respectively.  It is important to emphasize that these detectors are idealized and do not accurately take into account the limitations of the real experiments.  For example, Xenon experiments make use of time projection chamber detectors, which have poor resolution at low energies.  Thus, it is overly optimistic to assume that XENON1T has perfect resolution all the way down to the threshold, and properly taking into account the resolution at low energies will affect some of the near-threshold results in this paper.  On the other hand, Germanium detectors typically have excellent resolution all the way down to the threshold.} for 50, 100, and 500 GeV DM, assuming the SHM.  Roughly $10^4$ times more exposure is needed to observe the annual modulation over the constant rate for these three cases.  However, the annual modulation can be easier to see for DM that is: $(i)$ $O(10)$ GeV, $(ii)$ in a dark disk (DD) or stream (Str), or $(iii)$ inelastic with a mass splitting of $\delta=90$ keV.  In some of these cases, only $\sim10$ times more exposure is needed to see the annual modulation over the unmodulated rate. 

It is important to stress the significance of the enhancement of $E(A_0)/E(A_1)$ for the light DM and inelastic-scattering scenarios.  Consider an illustration of a more generic scenario.   Suppose that after $\sim 10$ days of running the LUX~\cite{Akerib:2012ys} experiment ($\sim$ 300 kg Xenon) detects an unmodulated rate from a $\sim 50$ GeV or heavier DM mass with a cross section $\sigma_0$ directly below the XENON100 limit~\cite{Aprile:2011dd,Aprile:2012nq}.  The scaling relation $E(A_0) / E(A_1) \sim \epsilon^4$ then indicates that a ton-scale detector like XENON1T cannot detect annual modulation within a reasonable timeframe.  In contrast, for light DM, $E(A_0) / E(A_1) \sim 10$ and can be observable, as we will discuss.

\begin{figure}[tb]
  \leavevmode
\begin{center}$
\begin{array}{cc}
\scalebox{.58}{\includegraphics{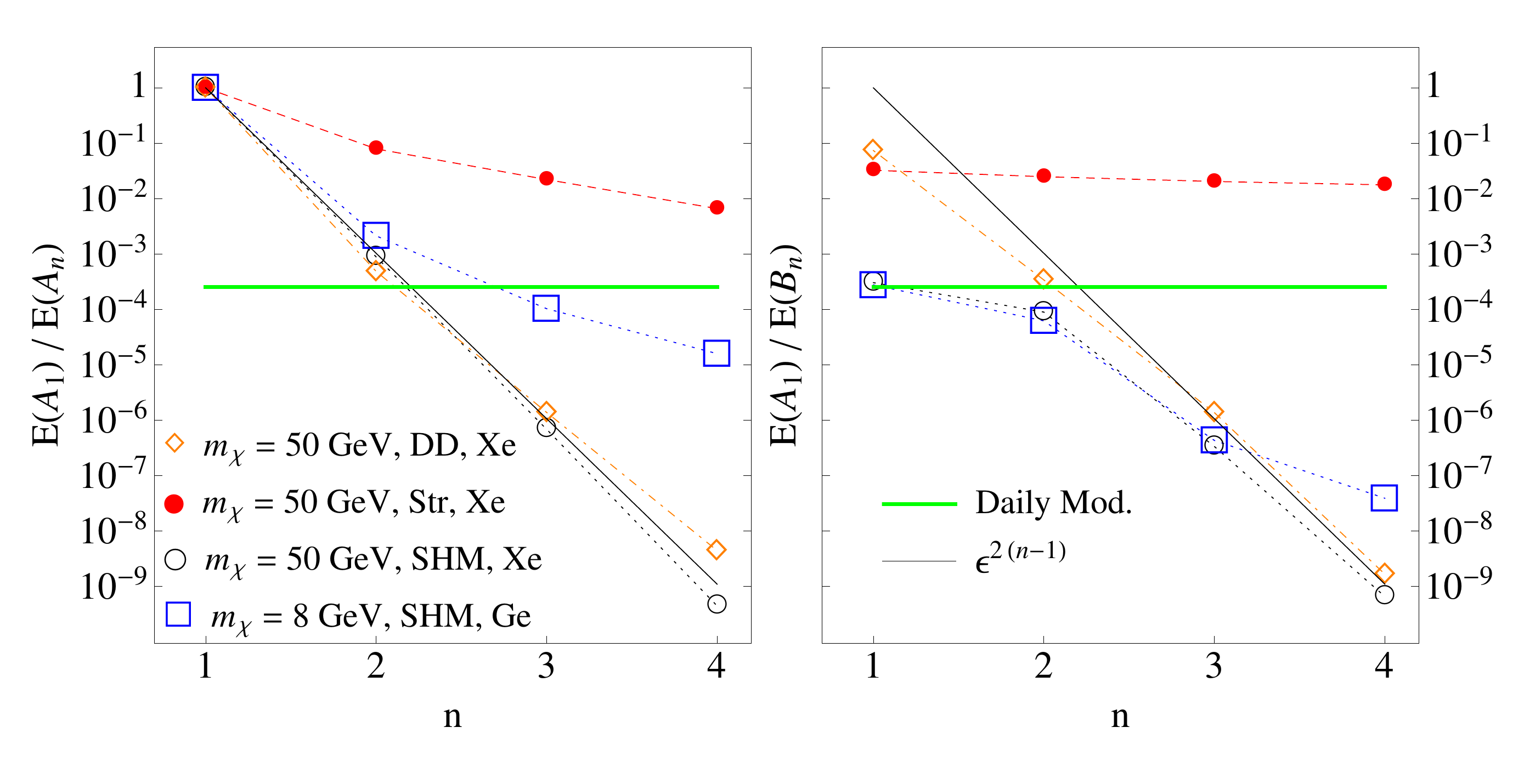}} 
  \end{array}$
\end{center}
\vspace{-1cm}
\caption{ The exposure needed to observe $A_1$ relative to that needed to observe $A_n$ (left) and $B_n$ (right), as a function of the mode $n$.  Generically, these ratios follow the scaling relations $E(A_1) / E(A_n) \sim  \epsilon^{2 \, (n-1)} $ and $E(A_1) / E(B_n) \sim  \epsilon^{2 \,(n-1)}$, with $E(A_1) / E(B_1) \sim \epsilon^{2}$ (black line).   The green horizontal line marks the value $E(A_1) / E(A_d) \approx 1 / 63^2$ for daily modulation at XENON1T assuming a standard isotropic distribution.  Note that ``Xe" denotes XENON1T and ``Ge" denotes GEODM DUSEL.  The dark disk (DD)  and stream (Str) configurations are the same as in Fig.~\ref{A0vA1}.
 }
\label{AnBnvA1}
\end{figure}

Figure~\ref{AnBnvA1} illustrates the dependence of $E(A_1)/E(A_n)$ (left) and $E(A_1)/E(B_n)$ (right) as a function of the mode number $n$.  The black line in both panels indicates the scaling relation $E(A_1) / E(A_n),E(A_1) / E(B_n)  \sim  \epsilon^{2 \, (n-1)} $.  This relation generally holds except for the modes $B_1$ and $A_d$, for which $E(A_1) / E(B_1) \sim E(A_1) / E(A_d) \sim  \epsilon^{2} $.  This scaling is clearly illustrated by the $m_\chi = 50$ GeV example assuming the pure SHM velocity distribution (black circle).  Halo substructure in the form of the dark disk (orange diamond) enhances the mode $B_1$.  The stream (red circle) has the most drastic effect on the higher-order harmonic modes, while the light DM scenario is similar (blue square).  

We will come back to the following three benchmarks over the course of this paper: DM with large threshold speeds (light or inelastic), dark disks, and streams. These benchmarks highlight instances where the higher-order harmonics are particularly pronounced and relevant in current and near-future experiments.  For example, consider the recent CDMS II Si~\cite{Agnese:2013rvf} result, which finds three events in the signal region, with a background expectation of 0.7.  This is consistent with a DM particle of mass 8.6 GeV and a cross section of $1.9 \times 10^{-41}$ cm$^2$.  If this light DM scenario is true, then GEODM DUSEL and XENON1T should see evidence for annual modulation and the second harmonic $A_2$ in the near future, as shown in Table~\ref{ResultsSHM}.  In particular, these experiments should observe the modes $A_1$ and $A_2$ in less than a year.  With a $2\kevnr$ threshold, GEODM DUSEL could also potentially observe $B_1$ and $A_d$ a few years later.  
\begin{table}
\begin{center}
\begin{tabular}{|  c |  cc c |}
\hline
Mode 	& {\bf XENON1T} 									& {\bf GEODM DUSEL } 	&		{\bf GEODM DUSEL}   	\\
   			& $E_\text{thresh}$: 4 keV$_\text{nr}$ 	&  5 keV$_\text{nr}$ 						& 		2 keV$_\text{nr}$  						\\   
\hline
$A_1$ 	&  $\leq$ 1 year 									&   $\leq$ 1 year 							& 		$\leq$ 1 year    \\
$A_2$ 	&  $\leq$ 1 year 									&   $\leq$ 1 year 							& 		$\leq$ 1 year  \\
$B_1$ 	& 	- 														&   - 												& 1 - 2 years \\
$B_2$ 	& - 														&   - 												& - \\
$A_d$ 	& - 														&  - 												& 2 - 3 years \\
\hline 
\end{tabular}
\end{center}
\caption{The predicted amount of time to detect the first few harmonic modes at XENON1T and GEODM DUSEL to 95\% confidence, assuming the SHM with $v_0 = 220 \kms$ and $v_\esc = 550 \kms$, a DM mass of 8.6 GeV, and a DM-nucleon cross section $\sigma_0 = 1.9 \times 10^{-41}$ cm$^2$.  Times greater than 3 years are not shown.
}
\label{ResultsSHM}
\end{table}

\section{Harmonic Analysis}    
\label{sec: mis}

Having developed some intuition for the relative strengths of the different harmonic modes, let us now focus on deriving explicit analytic expressions for them.  This requires knowing the velocity distribution in the lab frame, which is obtained by applying a Galilean boost to the distribution in the local DM rest frame $\tilde{f}(\textbf{v})$.  The velocities are shifted by 
\es{Eqvobs}{
\textbf{v}_{\text{obs}}(t)  = \textbf{v}_{\odot} + \textbf{V}_{\oplus}(t) + \V{V}_{(\phi_0,\lambda_0)}(t) \, ,
}
where $\textbf{v}_{\odot}$ is the velocity of the Sun's frame relative to the DM reference frame, $\textbf{V}_{\oplus}(t)$ is the Earth's velocity, and $\V{V}_{(\phi_0,\lambda_0)}(t)$ is the velocity of the experiment at the surface of the Earth, relative to the Earth's center.\footnote{We neglect the motion of the Earth about the barycenter of the Earth-Moon system, which contributes around order $\epsilon^3$ (similar to $A_3$ and $B_3$).}
When analyzing modes with frequency $O(\omega)$, the velocity $\V{V}_{(\phi_0,\lambda_0)}(t)$ is not important and may be dropped.  The following section considers daily modulation, in which case $ \V{V}_{(\phi_0,\lambda_0)}(t)$ is the crucial contribution and cannot be neglected.

The time dependence in the DM velocity distribution is due to the Earth's orbital motion, which is subdominant to the Sun's velocity by $O(\epsilon)$.  As a result, $\tilde{f}\left(\textbf{v} + \textbf{v}_{\text{obs}}(t)\right)$ may be Taylor expanded in $\epsilon$ and substituted into the expression for the mean inverse speed in~\eqref{recoilR} to obtain the Fourier series
\es{etaFourier}{
\eta(v_\Min,t) = a_0(v_\Min) + \sum_{n=1}^\infty \big[a_n(v_\Min) \cos n \omega (t - t_0) + b_n(v_\Min) \sin n \omega(t-t_0) \big] \,.
} 
This expression is easily related to that for the differential rate in~\eqref{drdEtyp}.  

To begin, we need an explicit expression for $\textbf{V}_{\oplus}(t)$ that allows us to expand the observed velocity $\textbf{v}_{\text{obs}}(t)$ in terms of $\epsilon$.   This is the subject of Sec.~\ref{sub: Velocities}.  Expansions for the mean inverse speed with distributions isotropic and anisotropic in the Galactic rest frame are derived in Sec.~\ref{sub: Isotropic} and~\ref{Sub: Anisotropic}, respectively. 

\subsection{Velocities} \label{sub: Velocities}

\begin{figure}[tb]
\leavevmode
\begin{center}$
\begin{array}{cc}
\scalebox{.5}{\includegraphics{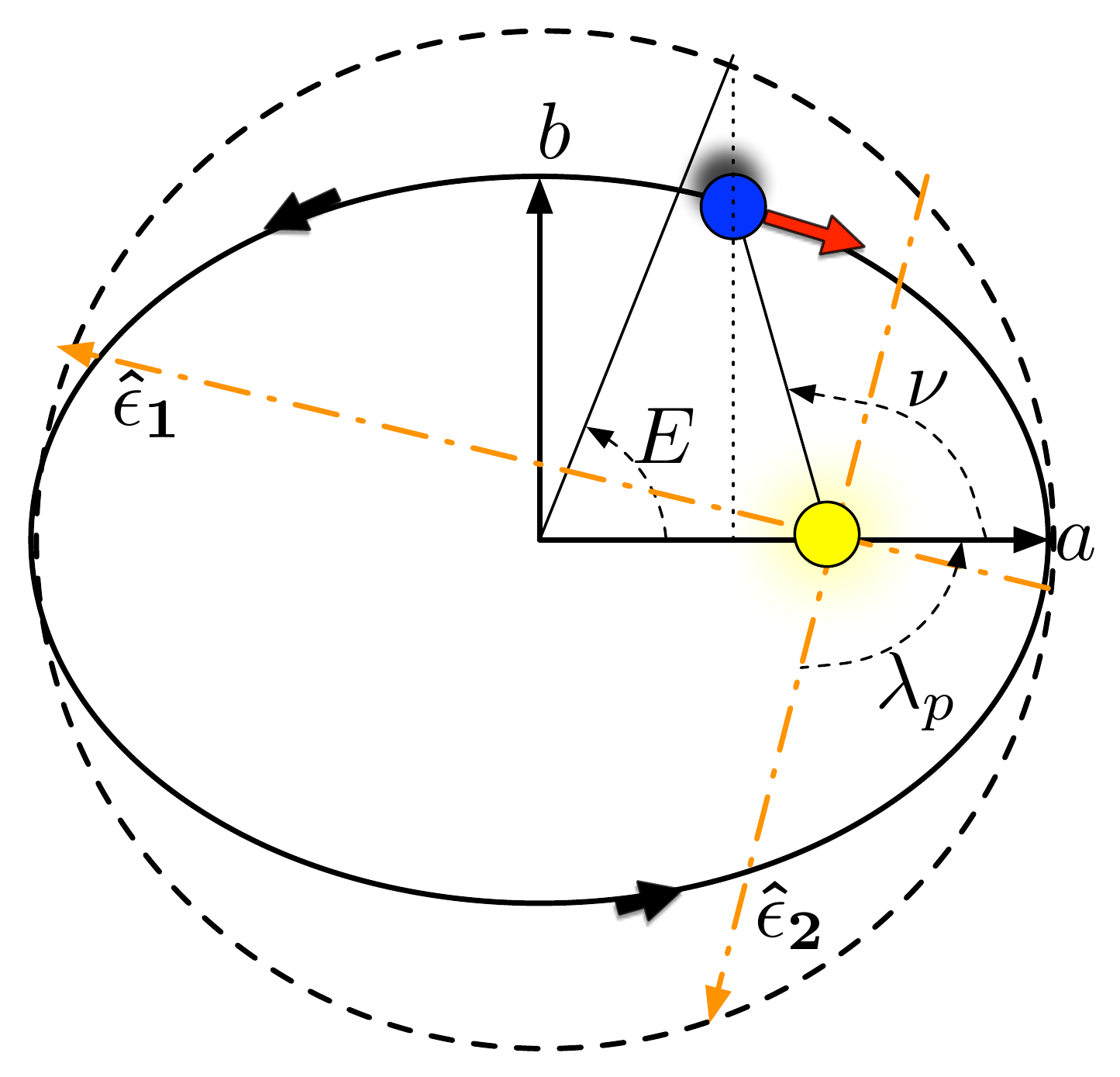}}
 \end{array}$
\end{center}
\caption{
The Earth (blue) orbits the Sun (yellow) counterclockwise along the solid-black ellipse.  The perihelion of the orbit is denoted by the letter $a$, which is also the length of the semi-major axis.  The Sun is located a distance $f = a \, e$ from the center of the ellipse, where $e$ is the eccentricity.  The eccentric anomaly $E$ is the angle between a fictitious point on the dashed-black circle of radius $a$, the center of the ellipse (circle), and the perihelion.  The true anomaly $\nu$ is the angle between the perihelion, Sun, and Earth.  At the vernal equinox, the vector $\V{\hat \epsilon_2}$ points from the Earth to the Sun.  Note that the Earth has just passed the vernal equinox in this diagram.  The ecliptic longitude of the perihelion $\lambda_p$ is the angle between the autumnal equinox, the Sun, and the perihelion.  The projection of the Earth's rotational axis to the ecliptic plane (red arrow) points in the $- \V{\hat \epsilon_1}$ direction.  The obliquity of the Earth's rotational axis $\varepsilon \approx 23^\circ.4$ is the angle between the Earth's rotational axis and $\V{\hat \epsilon_3}$ ($\V{\hat \epsilon_3} = \V{\hat \epsilon_1} \times \V{\hat \epsilon_2}$).  The diagram is not to scale.    }        
\label{EarthSun}
\end{figure}

For a generic choice of initial time $t_0$, $v_\text{obs}(t) \equiv | \V{v_\text{obs}}(t) |$ has the natural Fourier expansion
\es{CircExpand}{
v_\obs(t) = \bar v\left(1 + \sum_{n = 1}^\infty \epsilon^n v_n \cos n \omega (t - t_0) + \sum_{n = 1}^\infty \epsilon^n u_n \sin n \omega (t - t_0) \right) \,,
}
where by ``natural" we mean that the $v_n$ and $u_n$ harmonic coefficients remain order one at increasing $n$.  To solve for the expansion coefficients, we use an explicit expression for the Earth's velocity in terms of its orbital parameters, illustrated in Fig.~\ref{EarthSun}.  The most relevant parameters for our immediate purpose are the eccentricity of the orbit, $e\approx0.016722$, the ecliptic longitude of the perihelion, $\lambda_p\approx102^{\circ}$, and the unit vectors ${\bf \hat{\epsilon}_{1,2}}$ that span the ecliptic plane.   Appendix~\ref{sec: ORBIT} reviews the derivation of the general expression for the Earth's velocity in terms of its orbital parameters.  

To a good approximation, the velocity of the lab frame relative to the Galactic Center is 
\es{GoodExpand}{
v_\obs(t) \approx \bar v\big[1 + &\epsilon\, v_1 \cos \omega (t - t_0) \\
 &+ \epsilon^2 \big( u_1 \sin \omega (t - t_0) + v_2 \cos 2 \omega (t - t_0) + u_2 \sin 2 \omega (t - t_0) \big) \big]   \,,
}
with the harmonic coefficients given by
\es{vbarvnGood}{
\bar v &\approx v_\Sun \,, \qquad v_1 \approx 4 \,d\,, \qquad  v_2 \approx - 4\, d^2 -{4\, d \, e \over \epsilon} \cos(\lambda_p - \omega \, \phi) \,, \\
 u_1 &\approx{8\, d \, e \over \epsilon} \sin(\lambda_p - \omega \, \phi)   \,, \qquad u_2 \approx - {u_1/2} 
}
to leading order in $\epsilon$ ($e$), where
\es{phin}{
d  = \sqrt{( \V{\hat v}_\Sun \cdot \V{\hat \epsilon_1})^2 + ( \V{\hat v}_\Sun \cdot \V{\hat \epsilon_2})^2} \,, \qquad \phi = 
{1 \over \omega} \cos^{-1} \left( {\V{\hat v}_\Sun \cdot \V{\hat \epsilon_1} \over d} \right) \,.
}
The modes $\sin \omega (t - t_0)$  and $\sin 2 \omega (t - t_0)$ scale as $\epsilon \times e$.  This is a consequence of choosing $t_0 \approx t_1 + 73.4$ days, which is the time that maximizes $v_\obs(t)$ (see Appendix~\ref{sec: ORBIT} for more details), where $t_1$ is the time of the vernal equinox.  Because $e / \epsilon \approx 0.52$, the mode $\sin \omega (t - t_0)$ is order $\epsilon^2$ instead of order $\epsilon$, as would be the case for a more generic choice of $t_0$.  

The numerical values of the harmonic coefficients, which are used extensively throughout the rest of the paper, are:
\es{finalValues}{ 
\big\{ \bar v,\,v_1,\, v_2,\, u_1,\, u_2 \big\} \approx  \big\{ 232 \kms,\,1.96,\, -1.85,\, 1.04,\, - {u_1/ 2} \big\}.
}
Higher-order corrections to these leading values are straightforward to work out numerically, and they cause changes of at most $\sim 10\%$.  We note that Lewin and Smith~\cite{Lewin:1995rx} used an incorrect expression for $\V{V_\earth}(t)$ (see Appendix~\ref{sec: ORBIT} for more details), and using their velocity leads to errors on the order of $100$\% for $v_2$, $u_1$, and $u_2$.  Said another way, their expression for $\V{V_\earth}(t)$ leads to an erroneous $t_0$, which is off by around half a day.

\subsection{Distributions isotropic in the Galactic rest frame }\label{sub: Isotropic}

The Fourier coefficients $A_n$ and $B_n$ of the differential rate are directly proportional to the $a_n$ and $b_n$ coefficients (respectively) by the factor 
 \es{gammaDeff}{
   \Gamma( E_{\text{nr}} ) \equiv \sigma_0 {\rho_\chi A^2 \over 2 m_\chi \mu_p^2 } F^2(q) \, ,
  }
where $\mu_p$ is the DM-proton reduced mass, $A$ is the atomic number of the target nucleus, $\sigma_0$ is the spin-independent scattering cross section of the DM with protons and neutrons (assumed to be comparable), and $\rho_{\chi}$ is taken to be 0.4 GeV/cm$^3$~\cite{Caldwell:1981rj,Catena:2009mf,Weber:2009pt,Salucci:2010qr,Pato:2010yq}.  $F^2(q)$ is the Helm nuclear form factor~\cite{Smith:1988kw,Helm:1956zz}.
 \begin{figure}[tb]
  \leavevmode
\begin{center}$
\begin{array}{cc}
\scalebox{.59}{\includegraphics{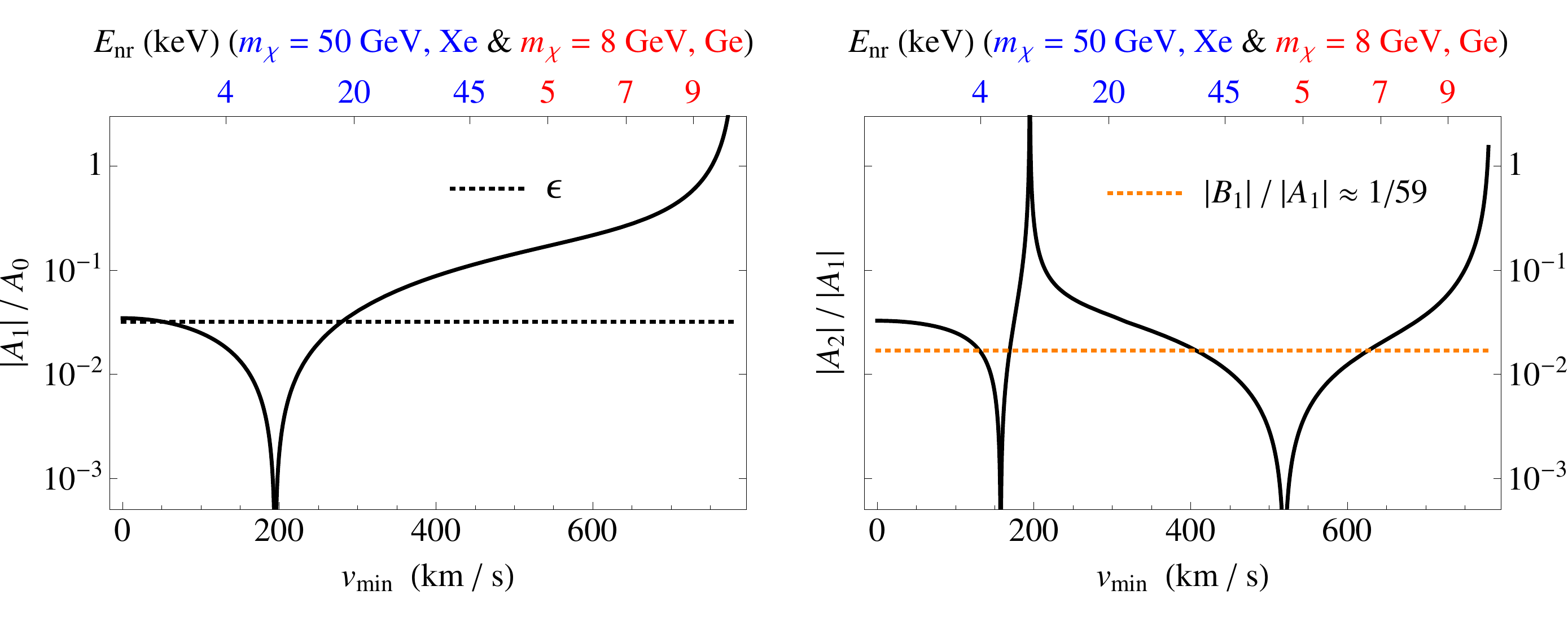}}
  \end{array}$
\end{center}
\vspace{-.5cm}
\caption{ The ratios of harmonic modes $|A_1| / A_0$ and $|A_2| / |A_1|$ within the SHM, which is taken to have $v_0 = 220 \kms$ and $v_\esc = 550 \kms$.  Note that the ratio $|B_{1}| / |A_1|$ is fixed~\eqref{abfractions} because the SHM is isotropic.  The $x$-axis may be shown either in terms of $v_\Min$ (bottom) or $E_\text{nr}$ (top).  In going from $v_\Min$ to $E_\text{nr}$, we need to know the scattering medium, whether the scattering is elastic or inelastic, and the DM mass.  We show values of $E_\text{nr}$ on the $x$-axis assuming elastic scattering in (a) Xenon, with $m_\chi = 50$ GeV (blue numbers) and (b) Germanium, with $m_\chi = 8$ GeV (red numbers).   
 }  
\label{abSHO}
\end{figure}
Note that $\sigma(q^2) = \sigma_0 (\mu^2 / \mu_p^2) \,F^2(q) \, A^2$.

In practice, when the velocity distribution is isotropic and smooth over scales $\sim V_\earth$, the $a_n$ and $b_n$ coefficients will decrease as powers of $\epsilon$ at increasing $n$.  A relatively straightforward calculation gives   
\es{anbnGEN}{
a_0(v_\Min) &\approx 2 \pi \int_{v_\Min}^\infty dv \, v \int_{-1}^{+1} ds \left[ \tilde f (v^2 + 2 s \,v \,\bar v + \bar v^2) \right]  \,, \\
a_1(v_\Min) &\approx 4 \pi \epsilon \, v_1 \, \bar v \int_{v_\Min}^\infty dv \, v \int_{-1}^{+1} ds (s\, v + \bar v) \tilde f' (v^2 + 2 s\, v\, \bar v + \bar v^2)  \,, \\
a_2 (v_\Min) &\approx { v_2 \epsilon \over  v_1} a_1(v_\Min) + \pi \epsilon^2 v_1^2  \bar v^2  \int_{v_\Min}^\infty dv \, v \int_{-1}^{+1} ds \left[ \tilde f' (v^2 + 2 s\, v\, \bar v + \bar v^2) \right. \\
&\left. +\, 2 (s\, v + \bar v)^2 \tilde f'' (v^2 + 2 s\, v\, \bar v + \bar v^2) \right] \,, \\
{b_1(v_\Min) \over a_1(v_\Min) } &\approx  2 \, e \, \sin (\lambda_p - \omega \, \phi) \,, \qquad {b_2(v_\Min) \over b_1(v_\Min) } \approx - {1 \over 2}  \,,
} 
to leading order in $\epsilon$.  Notice in particular that regardless of the form of the velocity distribution function $\tilde f$, so long as it is isotropic in the Galactic rest frame and smooth, 
\es{abfractions}{
{B_1(v_\Min) \over A_1 (v_\Min) } \approx {1 \over 59} \,, \qquad {B_2(v_\Min) \over B_1 (v_\Min) } \approx - {1 \over 2}  \,.
}
The fractions above should be contrasted to the fraction $A_2(v_\Min) / A_1(v_\Min)$, which depends non-trivially on $v_\Min$ even at leading order in $\epsilon$.   Even so, the magnitude of this ratio is set by the number 
\es{b1b2approx}{
{A_2(v_\Min) \over A_1 (v_\Min) } \sim { v_2 \, \epsilon \over  v_1} \approx - {1 \over 33} \,.
}
Further, it is interesting to note that because $u_2 \approx - (u_1 / 2)$, independent of astrophysical uncertainties such as errors in $\V{v}_\Sun$, one should find $B_2 / B_1 \approx - (1/2)$ for velocity distributions isotropic in the Galactic frame.  
Measuring this ratio is thus an excellent method to search for anisotropies in the DM velocity distribution.

Analytic expressions for the first few higher-harmonic coefficients for the Maxwell distribution and the SHM are given in Appendix~\ref{Examples}.  The analytic expressions are used instead of the full numerical results where possible throughout the paper.  We have verified that the analytic and numerical results are consistent.  

The left panel of Fig.~\ref{abSHO} shows $|A_1|/A_0$ as a function of $v_{\text{min}}$, assuming the SHM with $v_0 = 220 \kms$ and $v_\esc = 550 \kms$.  Our expectation is that this ratio scales as $\epsilon$, shown as the dotted line.  This is indeed the case, except in the region where $A_1$ has a zero ($\sim 195 \kms$).  The annual modulation mode is enhanced at larger threshold speed, becoming $O(1)$ near $v_{\text{min}} \sim 700 \kms$.  The right panel shows $|A_2|/|A_1|$, which from~\eqref{b1b2approx} should be $\sim 1/33$.  While this scaling is generally correct, the ratio is a non-trivial function of $v_\Min$.  The ratio $|B_1| / |A_1|$ (dotted orange), on the other hand, is constant as a function of $v_\Min$ (see~\eqref{abfractions}).   

Figure~\ref{abSHO} shows that the higher-order harmonics can have zeros at particular values of $v_{\text{min}}$.  The location of these zeros 
can be used to estimate the DM mass~\cite{Lewis:2003bv}.  For example, the modes $A_1$, $B_1$, $B_2$ all have approximately the same zero, which occurs at $v_\Min \approx 195 \kms$.\footnote{The daily-modulation mode $A_d$, which we discuss in the following section, also has the same zero.}  The mode $A_2$ has two zeros; the lower one occurring at $v_\Min \approx 163 \kms$ and the upper at $v_\Min \approx 554 \kms$.  The exact location of these zeros depends on assumptions about the astrophysical parameters.  Varying $v_0$ over $200 - 260 \kms$ and $v_\esc$ over $500 - 600 \kms$, the zero of $A_1$, $B_{1}$, and $B_2$ varies over the range $(180, 230) \kms$, while the zeros of $A_2$ vary over the ranges $(150, 200) \kms$ and $(500, 690) \kms$.         
 
\begin{figure}[tb]
  \leavevmode
\begin{center}$
\begin{array}{cc}
\scalebox{.58}{\includegraphics{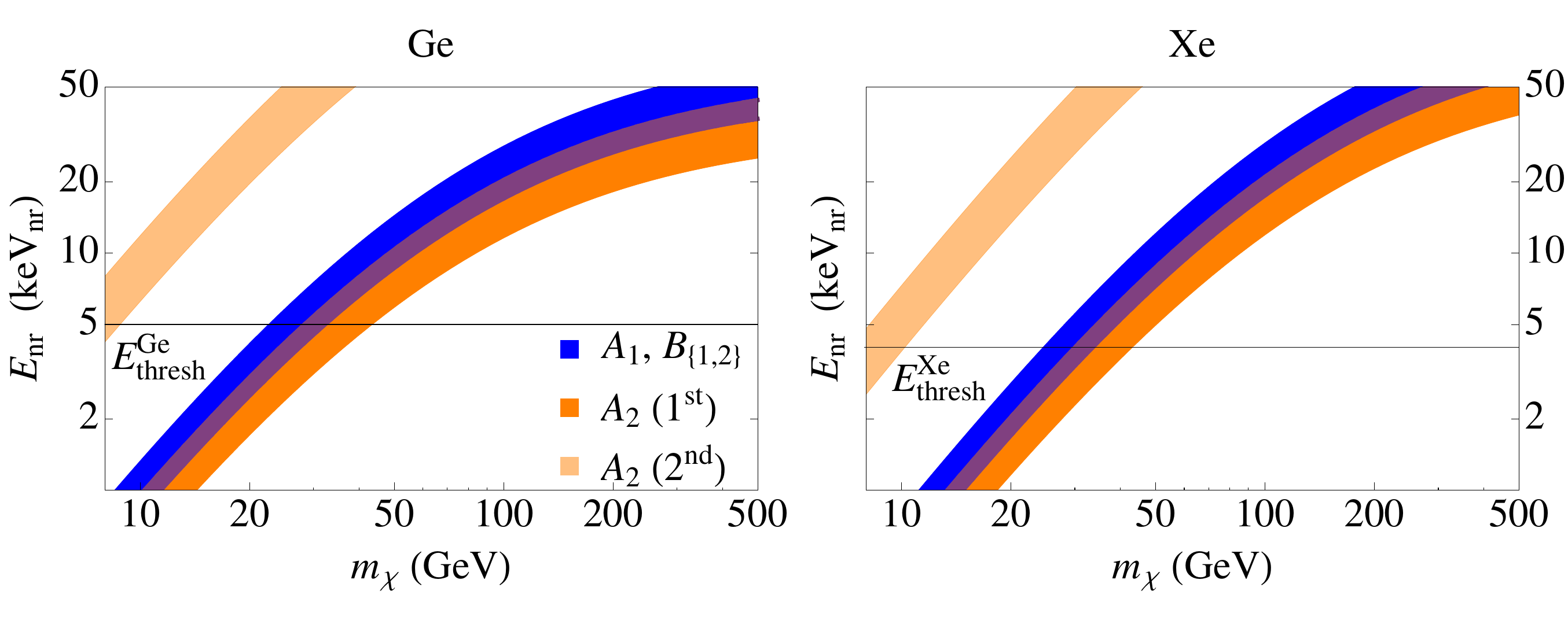}} \\
 \end{array}$
\end{center}
\vspace{-.5cm}
\caption{The values of $E_\text{nr}$ that zero the harmonic modes $A_{1,2}$, $B_{1,2}$, and $A_d$ as functions of the DM mass $m_\chi$ at Germanium (left) and Xenon (right) experiments, assuming elastic scattering.  The velocity distribution is taken to be the SHM.  Within this model $A_1$, $B_{1,2}$, and $A_d$ are proportional to each other, and these functions have one common zero.  The function $A_2$, on the other hand, has two zeros.  The shaded regions account for astrophysical uncertainties, varying $v_0$ over $200 - 260 \kms$ and $v_\text{esc}$ over $500 - 600 \kms$, with the purple region being the overlap of the blue $(A_1, B_{1,2})$ and dark orange $(A_2)$ regions.    }  
\label{determineMass}
\end{figure}

The plots in Fig.~\ref{determineMass} show the values of the recoil energy $E_\text{nr}$ that zero the various harmonic coefficients, for a given DM mass $m_\chi$.  We consider elastic scattering off of Germanium (left panel) and Xenon (right panel).  The colored regions indicate the spread from astrophysical uncertainties.  For DM masses $\lesssim 20$ GeV, the second zero of $A_2$ may be the only such zero that occurs at a value of $E_\text{nr}$ above the common threshold energies. 
  
\subsection{Distributions anisotropic in the Galactic rest frame } \label{Sub: Anisotropic}

If the velocity distribution $\tilde f(\V{v})$ is anisotropic within the Galactic rest frame, the results in~\eqref{anbnGEN} no longer hold and there will be deviations from the constant ratios in~\eqref{abfractions}.  Thus, measuring the ratios $B_1 / A_1$ and $B_2 / A_1$ as functions of $E_\text{nr}$ is one way to probe the anisotropy of the velocity distribution.
 
Anisotropic velocity distributions may be grouped into two categories.  The first consists of DM substructure with an isotropic velocity distribution in its own rest frame, which does not necessarily coincide with that of the Galaxy.  Two examples are dark disks and DM streams.  In this case, much of the machinery developed in Sec.~\ref{sub: Velocities} and \ref{sub: Isotropic} still holds, though the stream example is more subtle because the velocity distribution is not smooth over scales $\sim V_\earth$.   The second category consists of velocity distributions that are not isotropic in any inertial rest frame.  This scenario is not a straightforward generalization of the procedure presented in the previous two subsections, and a systematic study of the harmonic coefficients for these distributions is left for future work.   

Let us consider the case of substructure with an isotropic distribution in its own reference frame.  The results of Sec.~\ref{sub: Velocities} carry over in a relatively straightforward fashion, with the replacement $\V{v_\Sun} \to \V{v_\Sun^S}$, where $\V{v_\Sun^S}$ is the velocity of the Sun's rest frame as measured in the substructure's rest frame.  More specifically, the parameter $\epsilon$ is replaced by a new parameter $\epsilon_S$, defined by 
\es{epsilonS}{
\epsilon_S \equiv {V_\earth \over 4  v_\Sun^S } \,, \qquad v_\Sun^S = | \V{v_\Sun^S} | \,.
}
Perturbation theory in $\epsilon_S$ is valid so long as $ v_\Sun^S$ is sufficiently larger than $V_\earth$.  

When all of the local DM is in substructure, then the mean inverse speed has the same expansion as~(\ref{etaFourier}), except that $a_n, b_n \rightarrow a_n^S, b_n^S$ and $t_0 \rightarrow t_0^S$.
Similar to the SHM-like cases, where the DM distribution is isotropic in the Galactic rest frame, $a_1^S$ is naturally order $\epsilon_S$, while $b_1^S$, $a_2^S$, and $b_2^S$ are order $\epsilon_S^2$.  Indeed, these coefficients may be calculated perturbatively using equations~\eqref{anbnGEN} after making the appropriate substitutions.  In particular, the time $t_0^S$ at which the velocity of the DM in the lab frame is maximized can be quite different from that in distributions isotropic in the Galactic frame ($t_0$); a deviation from the expected $t_0$ could point to the presence of velocity substructure. 

When only a fraction of the local DM is in substructure, then we model the local distribution as having an additional SHM-like contribution.  In this case, it is natural to write the expansion of the mean inverse speed in terms of $\left(t - t_0\right)$ instead of $\left(t - t_0^S\right)$, since $\left(t - t_0\right)$ is the expected phase of the SHM-like contribution, which is isotropic in the Galactic frame.  Specifically, the following replacement is useful:
\es{invspeedrep}{
\cos n \omega \left(t-t_0^S\right) \rightarrow \cos n\omega \left(t-t_0 - \Delta t^S\right) \, 
}
and similarly for the $\sin n\omega(t-t_0^S)$ terms.  Using standard trigonometric identities, we can
absorb
the time difference $\Delta t^S$ into the harmonic coefficients.  This gives $a_0 = a_0^S$ and
\es{etaFourierF3}{
 \left( \begin{array}{c} 
 a_n \\
 b_n \\
 \end{array} \right) = R_S^n \,  \left( \begin{array}{c} 
 a_n^S \\
 b_n^S \\
 \end{array} \right) \,, \qquad
 R_S = \left( \begin{array}{cc}
 \cos \omega \Delta t^S & - \sin \omega \Delta t^S \\
 \sin \omega \Delta t^S & \cos \omega \Delta t^S \\
 \end{array} \right) \, .
} 
Because $a_1^S$ is typically much greater than $b_1^S$, it is possible to ignore the latter contribution to $a_1$ and $b_1$.  

One of the most important consequences of halo substructure is its effect on the mode $B_1$.  Remember that $B_1$ scales as $\epsilon^2$ for an isotropic distribution.  In the presence of substructure, its value can be enhanced.  In this case, $b_1 \approx a_1^S \sin\omega \Delta t^S$ (see~\eqref{etaFourierF3}), which is order $\epsilon_S$ as long as $\Delta t^S$ is not sufficiently small.  Moreover, if substructure dominates over the SHM-like contribution at a given $v_\Min$, then 
\es{A1B1F}{
{B_1 \over A_1} \approx \tan \omega \Delta t^S \,,
}
which, depending on the value of $\Delta t^S$, can be greater than the constant ratio (\emph{i.e.}, $1/59$) expected for distributions isotropic in the Galactic frame.  Therefore, an enhancement in the ratio of $B_1/A_1$ may indicate the presence of anisotropy.

\section{Daily Modulation}
\label{sec: dailymodulation}  

We now turn to the calculation of the daily modulation, which  arises from the rotation of the Earth about its axis.  Daily modulation is often discussed in the context of directional detectors that measure the recoil direction of a scattered nucleus, in addition to its energy~\cite{Spergel:1987kx}.  These experiments take advantage of the fact that the relative orientation of the detector with respect to the DM wind changes over the course of the day.  Specifically, the modulation is observed in the differential recoil rate at a particular angle.
  
The daily modulation that we refer to here is a different effect.  This modulation is the change in the differential rate (over all angles) due to the Earth's rotation; it has frequency $\omega_d$ and is due to $\V{V}_{(\phi_0,\lambda_0)}(t)$.  When considering the daily modulation, one may to a first approximation neglect the Earth's orbital motion $ \textbf{V}_{\oplus}(t) $. 

The daily-modulation mode depends on the location of the experiment on the surface of the Earth, given by its latitude $\phi_0$ and its longitude $\lambda_0$.
The latitude is normalized such that $\phi_0 = 0$ is the equator and $\phi_0 = \pi /2$ is the north pole.  Similarly, $\lambda_0$ takes values in the interval $[0, 2 \pi)$, and $\lambda_0 = 0$ is the longitude that is closest to the Sun at the vernal equinox -- \emph{i.e.}, at this longitude on the equator, the zenith points directly towards the Sun.  Increasing $\lambda_0$ moves one eastward around the Earth.    

For distributions isotropic in the Galactic frame, the daily modulation is captured by the $\omega_\text{d}$ frequency component of $v_\text{obs}(t)$.  Using the expression for the velocity of a point on the Earth's surface relative to its center, $\textbf{V}_{(\phi_0,\lambda_0)}(t)$, described in Appendix~\ref{sec: ROT}, one finds that 
\es{dailyvobs}{
v_\obs(t) \approx v_\Sun \left[ 1  - \cos \phi_0\, {V_d \over v_\Sun} d_d \cos \big( \omega_\text{d} (t - t_0^d) + \lambda_0  \big) + \cdots \right] \,,
}
where $d_d = \sqrt{ (\V{\hat v_\Sun} \cdot \V{ \hat \epsilon_2})^2 + \big[ (\V{\hat v_\Sun} \cdot \V{ \hat \epsilon_1}) \cos \varepsilon + (\V{\hat v_\Sun} \cdot \V{ \hat \epsilon_3}) \sin \varepsilon \big]^2 } \approx 0.68$.  Here, $\varepsilon$ is the obliquity of the Earth's rotational axis, and $t_0^d \approx t_1 +  2.9 \, \text{hours} $  (see Appendix~\ref{sec: ROT}), where $t_1$ is the time of the vernal equinox.
As already mentioned, $V_d / v_\Sun \approx 2 \times 10^{-3}$ is of a similar order of magnitude to $\epsilon^2 \approx 1 \times 10^{-3}$.  For this reason, the daily modulation is roughly of the same order as $A_2$, $B_1$, and $B_2$.

Over timescales of order a day, one may approximate the differential rate as in~\eqref{dailyMod}.  The mode $A_d$ is the daily-modulation mode, and a straightforward calculation shows that 
\es{dailyModRat}{
{A_d(v_\Min) \over A_1(v_\Min) } \approx - {\cos \phi_0 \over \epsilon \, v_1} {V_d \over v_\Sun} d_d \approx - {\cos \phi_0 \over 46} \,.
}
Importantly, this ratio is approximately constant as a function of $v_\Min$ (or equivalently $E_\text{nr}$).  
At GEODM DUSEL we then find $|A_d / A_1| \approx 1/64$
while at XENON1T $|A_d / A_1| \approx 1/63$.  Moreover, there should be a $117^\circ$ phase shift in the mode between the two experiments due to their difference in longitude.  For obvious reasons, an experiment at the south pole, such as DM-ICE~\cite{Cherwinka:2011ij}, would see no daily modulation.

\section{Examples}\label{sec: examples}

We now return to the three benchmarks introduced in Sec.~\ref{sec: Scaling}.  This section will focus, in particular, on DM with a large threshold speed, dark disks, and streams.  These examples illustrate cases where the annual and higher-frequency modes are enhanced relative to the typical scaling relation.  In some of these cases, only $\sim 10$ times more exposure is needed to observe annual modulation than the unmodulated rate, relative to the $\sim 10^4$ times more exposure that might have been assumed for a more generic scenario.  Similarly, in many enhanced scenarios, the higher-order modes are detectable soon after the detection of annual modulation.  An enhanced harmonic spectrum could therefore indicate that the DM has a large threshold velocity and/or is in substructure.

\subsection{Large Threshold Speed}

Dark-matter models sensitive to large $v_\Min$ exhibit enhanced harmonic modes.  The minimum speed for a nucleus to recoil with a given recoil energy is 
\es{vmin}{
 v_\Min = \sqrt{1 \over 2\, m_{n} \, E_\text{nr}} \left( {m_n E_\text{nr} \over \mu} + \delta \right) \, ,
 }  
where $m_n$ is the mass of the nucleus, and $\delta$ is the mass splitting between the initial and final DM states.  For elastic scattering, $\delta = 0$.  For inelastic scattering, the lighter DM state upscatters to the more massive one after interacting with the nucleus.  

As the DM mass $m_{\chi}$ decreases, the minimum scattering velocity increases.  This is because lower-mass DM needs to be traveling at a higher speed to give the same nuclear recoil energy as higher-mass DM.  As a result, DM with mass $O(10)$ GeV is more sensitive to the tail of the velocity distribution.  Because the tail of the distribution is less populated, a larger exposure is required to observe the unmodulated rate for low-mass DM. 
\begin{figure}[tb]
  \leavevmode
\begin{center}$
\begin{array}{cc}
\scalebox{.55}{\includegraphics{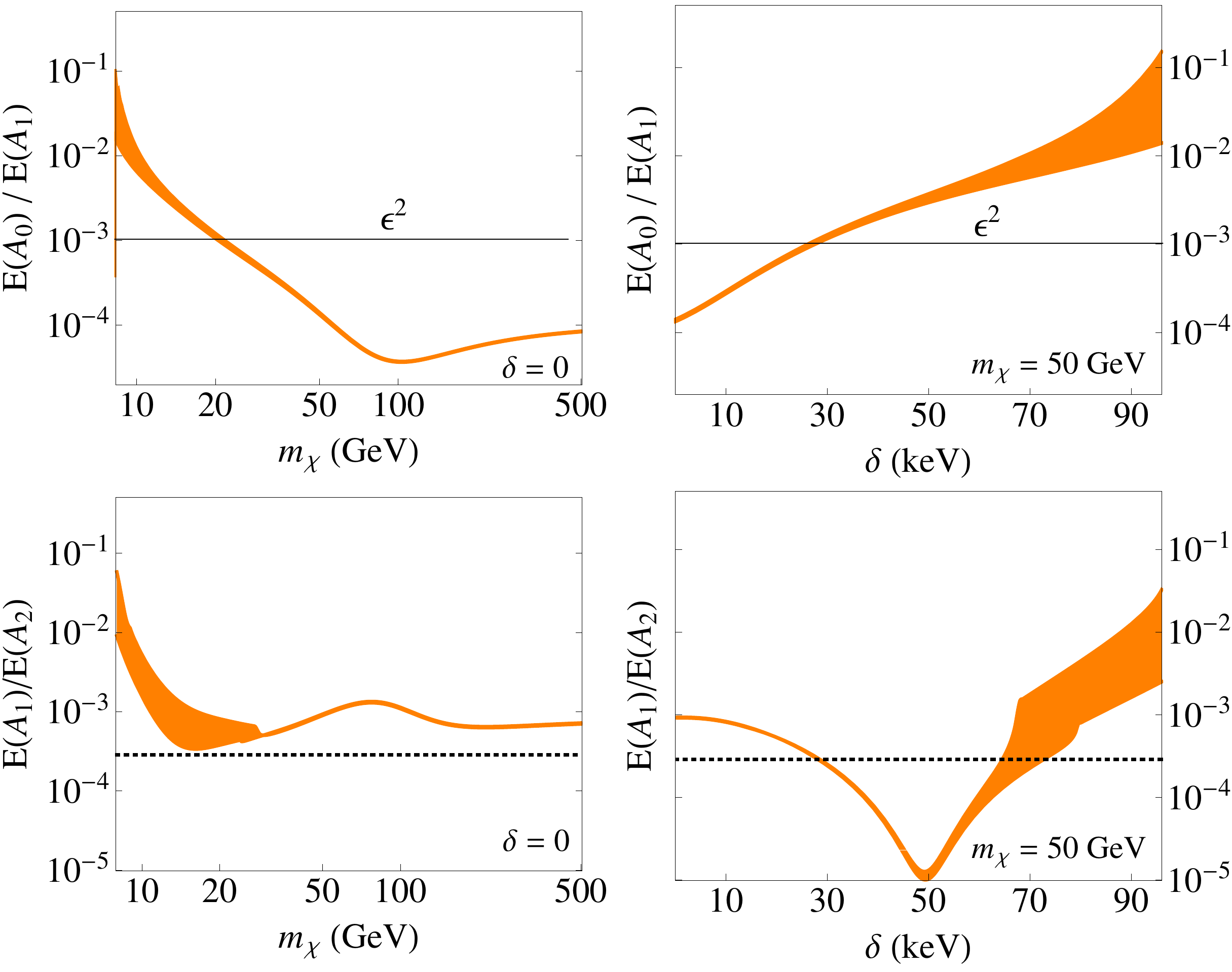}} \\
 \end{array}$
\end{center}
\caption{The left column illustrates how the ratios $E(A_0) / E(A_1)$ and $E(A_1) / E(A_2)$ depend on the DM mass in the elastic scattering scenario.  Decreasing the DM mass increases the relative magnitude of annual modulation and the higher-order harmonic modes.  A similar effect is seen in the inelastic-scattering scenario by increasing the mass-splitting $\delta$ (right column).  The pronounced dip in the lower-right plot is linked to the second zero of the mode $A_2$.  The dotted-black line in the bottom row marks $E(A_1) / E(B_1) \approx 1/59^2$ (see~\eqref{EA1B12}).  The orange shading indicates the spread from varying $v_\esc$ from $500 - 600$ km/s.  This figure is for XENON1T.       
}  
\label{imagesdm}
\end{figure}

Even though light DM gives fewer events in total, it enhances the significance of the annual modulation and the higher-order harmonics, as shown in Fig.~\ref{imagesdm}.  As one moves to lower DM masses (below $\sim20$ GeV), $A_1$ ($A_2$) is enhanced relative to $A_0$ ($A_1$).  The same is true for a fixed-mass DM (50 GeV) with increasing mass splitting $\delta$, because increasing the mass splitting increases $v_{\text{min}}$ as well. 

The behavior of the modes $B_1$, $B_2$ will be similar to that of $A_1$, shown in Fig.~\ref{imagesdm}, for an isotropic distribution.  The reason for this is that~\eqref{abfractions} implies  
 \es{EA1B12}{
{E(A_1) \over E(B_1)} \approx \left( {1 \over 59} \right)^2  \,, \qquad   {E(A_1) \over E(B_2)} \approx {1 \over 4} {E(A_1) \over E(B_1)} \,
}
to a good approximation.   
 
\subsection{The Dark Disk}  \label{DD}

N-body simulations find evidence for the formation of a dark disk in the Galactic plane of the Milky Way~\cite{Read:2008fh, Read:2009iv, Purcell:2009yp,Bruch:2008rx}.  The dark disk can form from the merger of massive satellites that are dragged into the baryonic disk and disrupted.  The ratio of the density of DM contained within the dark disk, $\rho_D$, to the density of the non-rotating halo at the solar position, $\rho_{\text{SHM}}$, is believed to be in the range $\rho_D / \rho_\text{SHM} \sim 0.5 - 2$~\cite{Read:2008fh, Read:2009iv}.  The dark disk rotates in the same direction as the Sun with a lag speed generally taken to be about $v_\text{lag} \approx 50 \kms$.  We model the velocity distribution of the dark disk as a Maxwellian with $v_0 \approx 70 \kms$ and assume that $\rho_{D} = \rho_{\text{SHM}}$~\cite{Bruch:2008rx}. 

\begin{figure}[tb]
\leavevmode
\begin{center}$
\begin{array}{cc}
\scalebox{.86}{\includegraphics{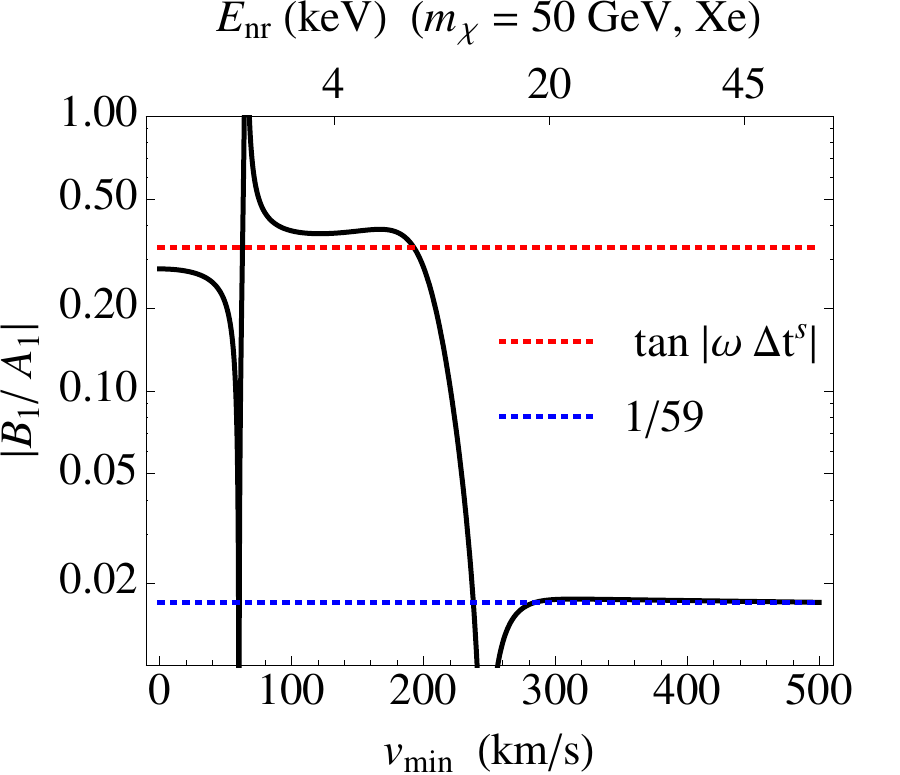}} & \scalebox{.82}{\includegraphics{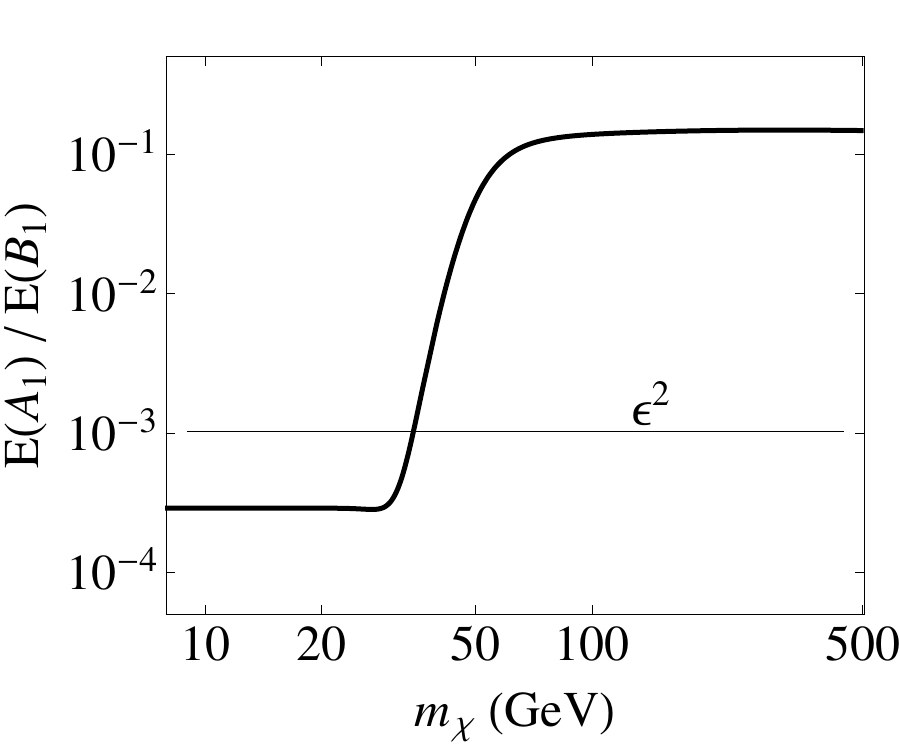}}   \\
 \end{array}$
\end{center}
\caption{
(left) The effect on the mode $B_1$ when the DM density is equally divided between the disk and SHM.  The ratio $|B_1 / A_1|$ (solid black) is significantly enhanced at low $v_\Min$ (or equivalently, low $E_\text{nr}$), where the dark disk velocity distribution is being probed.  The ratio asymptotes to the expected value for the dark disk at low velocities (dotted red) and to that for the SHM at high velocities (dashed blue).  (right)  The mode $B_1$ becomes more observable as the DM mass increases.
}  
\label{B1A1DD}
\end{figure}

The dark disk has the largest effect on the $B_1$ mode.  The ratio of $|B_1|/|A_1|$ is illustrated in the left panel of Fig.~\ref{B1A1DD} (solid black).  At large values of $v_\Min$, this ratio is approximately constant at a value of $1/59$ (dashed blue), which is consistent with what is expected for scattering off a distribution that is isotropic in the Galactic rest frame -- see \eqref{abfractions}.  At low $v_\Min$, the ratio asymptotes to $\tan | \omega \Delta t^S|$ (dotted orange) -- see \eqref{A1B1F}.  At large velocities, the SHM component dominates, while at low velocities, the dark-disk component provides the dominant contribution to the scattering.  For the dark-disk parameters taken here, $\sin |\omega \Delta t^S| \approx 0.34$ is sufficiently greater than zero that the enhancement of $B_1$ is quite substantial at low velocities (more than an order of magnitude).  The transition between the dark-disk--dominated region and the SHM-dominated region occurs for  $v_\Min\sim220 \kms$, which corresponds to nuclear recoil energies that are accessible at current detectors for relatively moderate to heavy DM masses.  For example, this transition point corresponds to $E_{\text{nr}}\sim10$ keV for a 50 GeV DM in a Xe experiment.  The right panel of Fig.~\ref{B1A1DD} shows that $B_1$ is enhanced relative to $A_1$ for DM masses greater than $\sim40$ GeV.  This is because heavier DM corresponds to lower $v_{\text{min}}$ and is therefore sensitive to the bulk of the velocity distribution, not just the high-velocity tail.

Because $v^S_\Sun \approx 50 \kms$ is the same magnitude as $V_\earth$, one might worry that perturbation theory to leading order in $\epsilon_S$ is not sufficiently accurate.  To test the expansion, we compute $\Delta t^S$ using the approximations in~\eqref{t0approxF} and~\eqref{udvd} and find $\Delta t^S \approx - 25$ days.  A more accurate, numerical calculation using the exact expression for $\V{V_\earth}(t)$ gives $\Delta t^S \approx - 20$ days.  We conclude that the leading terms in $\epsilon_S$ are sufficient to qualitatively capture the effects of the dark disk, with deviations from the true values expected to be around 25\%.

\subsection{Dark-Matter Streams}
  
  The higher-order harmonic coefficients may also be enhanced if the velocity distribution is not smooth, such as for DM streams.  A DM stream forms from debris that is tidally stripped from an in-falling subhalo that has not yet had time to virialize~\cite{Diemand:2006ik,Freese:2003tt}.  Streams are characterized by their negligible velocity dispersion, and their velocity distribution can be modeled as a delta function, $\tilde f(\textbf{v}) = \delta^3(\textbf{v})$, within the DM rest frame~\cite{Freese:2012xd}.  Even if the density of the stream is only a small fraction of the density of the smooth component of the halo, it can dramatically increase the magnitude of the higher-order harmonic coefficients in the differential rate.  Figure~\ref{AnMod} shows why this is the case; the scattering rate for DM in a stream can be ``on" for part of the year and ``off" for the rest.  The Fourier decomposition of the differential rate then decays weakly with increasing harmonic number just as in the Fourier series of the box function.  If the scaling $A_n / A_0 \sim \epsilon^n$ and $B_n / A_0 \sim \epsilon^n$ is not observed, and instead the harmonic coefficients appear to be significantly enhanced at increasing $n$, this would provide strong evidence for a stream.    
  
To begin, we consider the scenario where a component of the DM is in a stream, modeled as a delta function, and another component is in the SHM.  The ratio of the densities in either component is taken to be $\rho_{\text{Str}} = 0.1\rho{_{\text{SHM}}}$.  The harmonic modes within the pure SHM are suppressed by factors of $\epsilon^n$, while in the combined SHM + stream model, they are suppressed by the ratio of the stream density to the SHM density, $\rho_\text{Str} / \rho_\text{SHM}$, for nuclear recoil energies in the range that probes the stream's velocity distribution.  
More specifically, when $v_\Min$ is in the narrow range\footnote{The variables $\bar v^S$ and $v_1^S$ are defined in an analogous way to $\bar v$ and $v_1$.} 
\es{streamGoodRange}{
\bar v^S (1 - \epsilon_S | v_1^S |) < v_\Min < \bar v^S(1 + \epsilon_S |v_1^S|) \,,
}
 the coefficients $a_n^S$ are approximated by
\es{anFinStream}{
a_n^S(v_\Min) \approx {2 \over n \, \pi \, \bar v^S} \sin \left[ n  \cos^{-1} \left( {v_\Min - \bar v^S \over \epsilon_S \, v_1^S \, \bar v^S} \right)  \right] \,\text{sign} (v_1^S) \,,
}
while the $b_n^S$ are suppressed relative to the $a_n^S$ by terms of order $\epsilon_S$ (see Appendix~\ref{Examples} for details).  The $a_n$ and $b_n$ coefficients are then related to $a_n^S$ and $b_n^S$ through~\eqref{etaFourierF3}.
\begin{figure}[tb]
\leavevmode
\begin{center}$
\begin{array}{c}
\scalebox{0.58}{\includegraphics{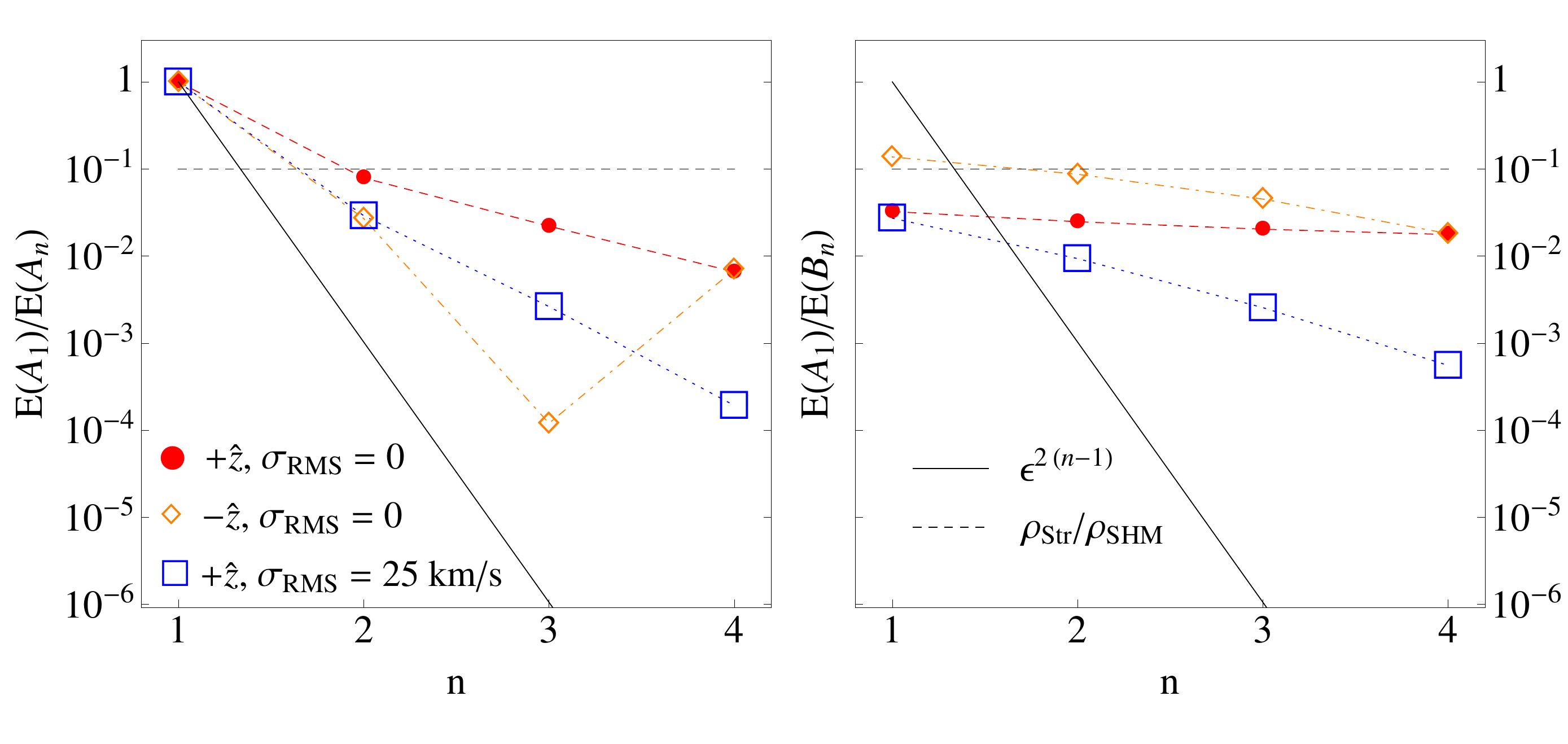}} \\
 \end{array}$
\end{center}
\vspace{-1cm}
\caption{   
Dark-matter streams significantly enhance the higher-order harmonic modes relative to the typical scaling relations.  In these examples, we have added the stream velocity distribution to that of the SHM, taking $\rho_\text{str} / \rho_\text{SHM} = 0.1$; the density ratio sets the scale for the quantities $E(A_1) / E(A_n)$ and $E(A_1) / E(B_n)$.  We consider two scenarios at a Xenon detector, both with $m_\chi = 50$ GeV: $(i)$ the stream travels in the $+{\bf \hat z}$-direction at a speed of $350 \kms$, and $(ii)$ the stream travels at the same speed in the $-{\bf \hat z}$-direction.  The $a_3$ mode is suppressed for the  $-{\bf \hat z}$ stream because, in that case, $\omega \Delta t^S \approx 5\, \pi / 6$.  Adding a small dispersion $\sigma_\text{RMS}$ to the stream does not significantly change the general behavior.   
 }  
\label{barStream}
\end{figure} 

Figure~\ref{barStream} illustrates $E(A_1)/E(A_n)$ (left panel) and $E(A_1)/E(B_n)$ (right panel) for 50 GeV DM at a Xe detector, for three different stream examples.  A dispersion-less stream that travels in the $+ {\bf \hat z}$ direction at $350 \kms$ leads to the most enhanced values of $A_n$ (red circle) of the three cases considered.  This enhancement is reduced if one treats the stream's velocity distribution as Maxwellian with small $\sigma_\text{RMS} =  25 \kms$ (blue square).  A dispersion-less stream with the same speed, but oriented in the opposite direction (orange diamond), has the most enhanced $B_n$ values of the three.

After applying the rotation matrix $R_S$, the $a_n$ and $b_n$ coefficients are generally of the same order.  However, interesting phase structure may emerge if $\Delta t^S$ takes special values.  For example, if $\Delta t^S \approx 0$, then one simply finds $b_n = b_n^S$ and $a_n = a_n^S$, which implies that the $b_n$ coefficients are suppressed relative to the $a_n$ coefficients.  
For the stream traveling in the $- \V{\hat z}$ direction,  $\omega \Delta t^S \approx 5\, \pi / 6$, which implies that $b_3 \approx a_3^S$, with $a_3^S$ suppressed.  This suppression in $A_3$ is apparent in Fig.~\ref{barStream} and nicely illustrates how the spectrum of harmonic modes provides a unique signature of the underlying astrophysical assumptions going into the DM model.

\section{Discussion} \label{sec: conclusions} 

The scattering rate at a direct-detection experiment modulates annually due to the motion of the Earth around the Sun and can deviate from a perfect sinusoid depending on the particulars of the DM model.  While annual modulation is a key feature of the spectrum, contributions from higher-order Fourier modes can provide additional information; for example, the zeros of the higher-order modes may be used to estimate the DM mass.  We have shown that the Fourier modes of the DM signal are related simply and elegantly to the parameters of the Earth's orbit.  For velocity distributions isotropic in the Galactic frame, certain ratios of these modes are constant; a measurement that shows a deviation from a constant value might indicate the presence of anisotropy in the halo.  

A simple scaling relation exists for the relative strengths of the higher-harmonic modes.  Specifically, the modes scale as $\epsilon^n$, where $\epsilon\sim 0.032$ for a standard isotropic distribution whose frame coincides with that of the Galaxy.  The mode $B_1$ is an exception, and is further suppressed by the eccentricity of the Earth's orbit.  The daily modulation in the total rate is roughly $O(\epsilon^2)$, similar to $A_2, B_1$ and $B_2$.  Note that our analysis requires an accurate model of the Earth's elliptical motion; we derived an accurate expression for the Earth's velocity  $\V{V}_\earth(t)$, which corrected an error in~\cite{Lewin:1995rx}.  The phase of the modulation obtained using the incorrect $\V{V}_\earth(t)$, for example, is off by about half a day from our value.     

Assuming 50 GeV DM and the SHM, $\sim10^4$ times more exposure is needed to observe an annual modulation (with 95\% confidence) after the unmodulated rate is detected.  An additional $\sim 10^3$ times more exposure is then needed to observe the second-order harmonic modes to 95\% confidence.  Given the current limits, observing any modulation whatsoever for DM heavier than $\sim 50$ GeV is challenging for ton-scale detectors.

However, annual and higher-order modes are enhanced for certain astrophysical and particle-physics scenarios.  The examples we consider here are light DM, inelastic scattering, and halo substructure, such as a dark disk or stream.  Light $O(10)$ GeV DM tends to strongly enhance the annual modulation as well as the higher-order modes (inelastic scattering is similar).  Substructure results in a modest enhancement in the annual modulation, but strongly affects the higher-order modes -- more so, even, than light DM.  These enhancements in either or both the annual and higher-order modes are highly relevant for both light DM and streams.  The dark disk has the largest effect on the mode $B_1$, but this is challenging to observe because it is most relevant for $\gtrsim 50$ GeV DM, which will be difficult to see modulating at a ton-scale detector.

There are currently dozens of direct-detection experiments in operation around the world, with even more planned to come into operation in the next few years.  Some of these experiments are already sensitive to these higher-order harmonics, while ton-scale experiments should definitely be able to provide confirmation of potential signals.  Let us consider some potential detections here, to illustrate the applicability of the higher-harmonic analysis.  Note that there is considerable debate as to whether these signals are due to DM given null results from~\cite{Aprile:2010bt,Angle:2007uj,Angle:2011th,Aprile:2011dd,Aprile:2012nq, Akerib:2005zy,Ahmed:2009zw,Ahmed:2010wy,Ahmed:2012vq,Ahmed:2011gh,Armengaud:2012pfa}.  A detailed discussion of how to reconcile these apparently conflicting results is beyond the scope of this work. 

The DAMA experiment~\cite{Bernabei:2008yh} consists of 250 kg of NaI(Tl) and claims an annual modulation with significance around 9$\sigma$~\cite{Bernabei:2010mq}.  If the scattering is predominantly off of sodium, the signal can be interpreted as coming from a light DM with mass near $10$ GeV~\cite{Bottino:2003iu,Bottino:2003cz,Gondolo:2005hh,Petriello:2008jj,Chang:2008xa,Savage:2008er}.  If this interpretation is correct, it would take DAMA $O(50)$ more years to observe the modes $A_2, B_1, B_2$, and $A_d$ to 95\% confidence.  While this is not feasible, the current signal can already be useful in limiting other scenarios.  For example, DAMA has presented a power spectrum of their signal (see Fig. 2 in~\cite{Bernabei:2010mq}), which limits $|A_2/A_1| \lesssim 0.2$.  This is already in tension with the benchmark stream considered in this paper (\emph{i.e.}, 350 km/s pointing in the $\hat{z}$ direction with $\rho_\text{Str} = 0.1 \rho_\text{SHM}$) for 50 GeV DM.
\begin{figure}[tb]
\leavevmode
\begin{center}$
\begin{array}{cc}
\scalebox{.58}{\includegraphics{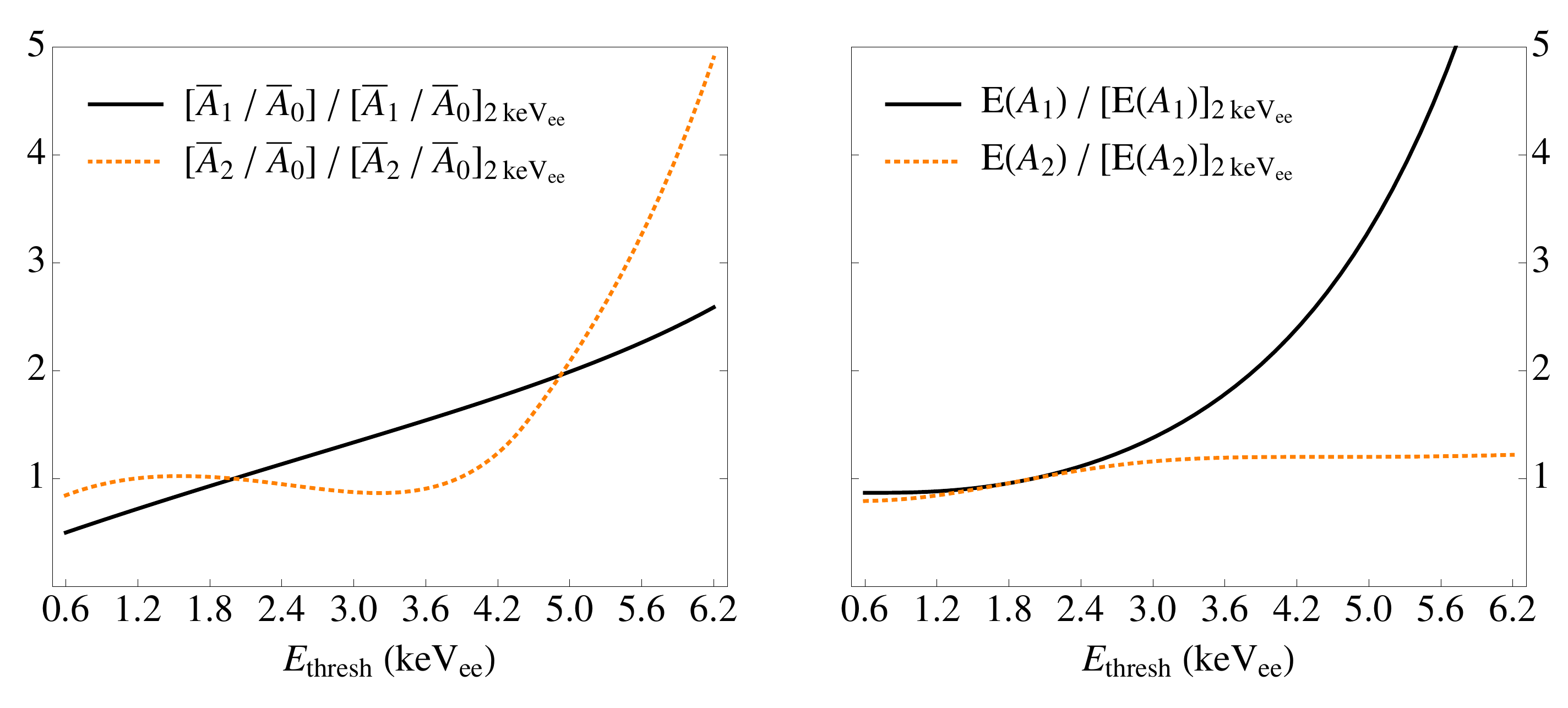}} \\
 \end{array}$
\end{center}
\vspace{-.75cm}
\caption{ The modulation fraction $\bar A_1 / \bar A_0$ and $\bar A_2 / \bar A_0$ (left panel), normalized with respect to their values at $E_\text{thresh} = 2$ keV$_\text{ee}$, as functions of the threshold energy $E_\text{thresh}$ for $11$ GeV DM scattering off of Na (relevant for the DAMA experiment).  Increasing $E_\text{thresh}$ causes $\bar A_1 / \bar A_0$ to increase, while $\bar A_2 / \bar A_0$ is less sensitive below $\sim 4.2$ keV$_\text{ee}$.  The right panel illustrates a related point, which is that increasing the threshold makes the discovery of annual modulation more difficult, while the mode $A_2$ is less sensitive to such changes.  The physical reason for this is that $A_2$ receives its dominant contribution from high $v_\Min$, which corresponds to $E_\text{nr}$ greater than the threshold value.  }
\label{imagesForEnd}
\end{figure}  

There is even further information to glean from the DAMA modulation.  As Fig.~\ref{imagesForEnd} illustrates, the modulation fraction has a distinctive shape as the threshold energy is varied.  If we define $\bar{A}_n = \int_{E_{\text{thresh}}}^{E_{\text{max}}} |A_n| \,dE_{\text{nr}} $, then the modulation fraction is $\bar{A}_1/\bar{A}_0$ (left panel, solid black).  This increases in a distinctive fashion as the lower threshold is increased.  (Note that the modulation fraction is normalized relative to that with a 2 keV$_\text{ee}$ threshold).  Similar behavior is observed for $\bar{A}_2/\bar{A}_1$ (left panel, dotted orange), although the shape of the curve is different -- it is nearly constant at low threshold energy and then increases sharply.  The right panel in Fig.~\ref{imagesForEnd} shows that 
increasing the threshold by a small amount may significantly increase the required exposure to detect annual modulation.  However, the mode $A_2$ is less sensitive to the threshold because that mode receives its dominant contribution from high $v_\Min$, which corresponds to $E_\text{nr}$ well above the threshold.  While the $A_2$ mode cannot be observed in a realistic timescale at DAMA, we discuss this mode here because a similar trend is also observed for other experiments.

The CoGeNT experiment~\cite{Aalseth:2010vx,Aalseth:2011wp,Aalseth:2012if} also claims to see an annually modulating signal with around $2.8\sigma$ significance, which can be interpreted in terms of DM~\cite{Frandsen:2011ts,Hooper:2011hd,Fox:2011px}.  CoGeNT needs $O(10^3)$ times more exposure to see the second-order harmonics; because CoGeNT is $\sim 0.4$~kg, it can never detect these modes.  However, a ton-scale Ge detector, such as GEODM DUSEL, would achieve the necessary exposure in a year or so.  To see our benchmark stream example, CoGeNT would need only $\sim15$ times more exposure. 

As illustrated in Table~\ref{ResultsSHM} for the CDMS II-Si excess~\cite{Agnese:2013rvf}, the ton-scale detectors XENON1T and GEODM DUSEL are particularly sensitive to these higher-harmonic modes.  This is because these experiments gain a marked increase in exposure due to their size, as opposed to integration in time.  These experiments could potentially confirm the CDMS II-Si result by observing the annual and higher-frequency modes in a few years, as discussed in Sec. II.

An annually modulating signal has long been considered critical for testing the DM hypothesis.  In cases where the annual modulation is observable by current and ton-scale detectors, higher-order harmonics are also enhanced.  This work only just begins to explore the various astrophysical effects that influence the harmonic structure, focusing specifically on the eccentricity of the Earth's orbit and simple examples of the velocity distribution and substructure.  In~\cite{Lee:2013wza}, we discuss the effects of gravitational focusing by the Sun and Earth, which may be comparable (see, for example,~\cite{Alenazi:2006wu}).  More generally, such astrophysical effects, combined with the particle-physics properties of the DM, leave a unique fingerprint on the harmonic spectrum that can be used to piece together clues about the missing matter.

\section*{Acknowledgments}
We thank E.~Figueroa-Feliciano, A. Green, and D. Spergel for helpful discussions.  We also thank A. Peter for cross-checking our derivation of the Earth's velocity.  ML is supported by a Simons Postdoctoral Research Fellowship.  This research was supported in part by the NSF grant PHY11-25915.  BRS was supported in part by the NSF grant PHY-0756966.

 \appendix
 
  \section{Statistical Analysis}  \label{poisson}

This section lays out the general statistical formalism required to answer the question: \emph{What is the typical amount of data necessary to detect a specific harmonic mode at a desired statistical significance, for a given set of DM model and experimental parameters?}  We apply the formalism presented here throughout this paper; for example, see Table~\ref{ResultsSHM}, in which we show the amount of time required to detect the first few harmonic modes at ton-scale experiments to 95\% confidence, assuming the SHM with typical parameters and WIMP particle properties consistent with the CDMS II Si result.

Our approach uses the results derived in \cite{Cowan:2010js}, which discusses the use of a typical, representative ``Asimov'' data set in likelihood-based tests for new physics.  The use of this typical data set allows one to make statements about the \emph{expected} discovery or exclusion significance of an experimental scenario (the significance realized by actual data will of course be random).  In particular, our final result is a straightforward extension of the statistical criterion presented in that paper for discovering a signal over background; in our case, the ``signal'' of interest is a specified harmonic mode and the ``background'' is the unmodulated event rate.

We begin by constructing the Asimov data set, for a given set of DM and experimental parameters.  We then apply a likelihood-ratio test to this data to derive a lower limit on the typical exposure needed to reject the null hypothesis of an unmodulated event rate and hence claim discovery of the specified mode.  To simplify the discussion, we focus on the first-order modulation; the results easily generalize to the higher-order modes.

The Asimov data set is the data set given by the mean event rate expected at the experiment.  Consider an experiment with detector mass $M$ that counts a number of events $N_{ij}$ in energy bins of width $\Delta E_\text{nr}$ centered around energies $E_{\text{nr},i}$ and time bins of width $\Delta t$ centered around times $t_j$.  Given a set of fixed model parameters $\boldsymbol\theta$ (which may include parameters for the experiment specifications, velocity distribution, DM properties, etc.), the mean number of events expected in each bin is
\begin{equation} \label{meansignal}
\lambda_{ij}(\xi; \boldsymbol\theta) = M \Delta E_\text{nr} \Delta t \left[A_0(E_{\text{nr},i}; \boldsymbol\theta) + \xi A_1(E_{\text{nr},i}; \boldsymbol\theta) \cos \omega (t_j-t_0) \right]\,.
\end{equation}
The coefficients $A_n(E_\text{nr}; \boldsymbol\theta)$, defined in~\eqref{drdEtyp}, are evaluated for the assumed model parameters.  The single free parameter $\xi$ determines the strength of the modulation signal.  Note that $\xi = 0$ corresponds to the null hypothesis of a background-only unmodulated event rate, while $\xi = 1$ corresponds to the alternative hypothesis of a modulation signal present at the level $A_1(E_\text{nr}; \boldsymbol\theta)$ predicted by the model.  The Asimov data set is given by the latter case,
\begin{equation} \label{asimovsignal}
\lambda_{ij,A} \equiv \lambda_{ij}(\xi = 1; \boldsymbol\theta)\,.
\end{equation}
From here on, we suppress the dependence on the model parameters $\boldsymbol\theta$.

Assuming Poissonian statistics, the likelihood function for the modulation-strength parameter $\xi$, given the Asimov data set $\lambda_{ij,A}$, is then
\begin{equation} \label{lfpoisson}
\mathcal{L}_A(\xi) = \prod_{ij} \frac{e^{-\lambda_{ij}(\xi)} \left[\lambda_{ij}(\xi)\right]^{\lambda_{ij,A}}}{\lambda_{ij,A}!}\,.
\end{equation}
The logarithm of the likelihood function can be Taylor expanded around the maximum-likelihood parameter value $\hat{\xi}$ that solves ${\partial \mathcal{L}_A(\xi)/\partial\xi|_{\xi = \hat\xi} = 0}$.  Using \eqref{meansignal}--\eqref{lfpoisson} leads to $\hat{\xi} = 1$ as expected, since this is the value of the modulation strength that was used to generate the Asimov data set.  The Taylor expansion around this value is approximately a Gaussian likelihood
\begin{equation} \label{lfxigauss}
\mathcal{L}_{A,g}(\xi) \propto \exp\left[-\frac{(\xi-1)^2}{2\sigma_A^2}\right]\,,
\end{equation}
where
\es{sigmaxi}{
\sigma_A^{-2} &= -\frac{d^2 \ln \mathcal{L}_A(\xi)}{d\xi^2}\bigg|_{\xi=1}\\
&\approx M \Delta E_\text{nr} \Delta t \sum_{ij} \frac{A_1^2(E_{\text{nr},i}) \cos^2 \omega (t_j-t_0)}{A_0(E_{\text{nr},i})}\\
&\approx \frac{M T}{2} \int_{E_\text{thresh}}^{E_\text{max}}\! dE_\text{nr}\, \frac{A_1^2(E_\text{nr})}{A_0(E_\text{nr})}\,
}
is found using \eqref{meansignal}--\eqref{lfpoisson}. 
  Two successive approximations have been made here.  First, we neglect terms that are higher order in $A_1/A_0$.\footnote{In most of the examples we consider, $E(A_1) \gg E(A_0)$, $E(A_n) \gg E(A_1)$, and $E(B_n) \gg E(A_1)$ ($n >1$).  It should be noted, however, that when these approximations break down, the Gaussian approximation may not give an accurate estimate for the required exposure.}  Second, we assume that the bin sizes and frequency $\omega$ are such that the discrete sums can be replaced with their corresponding integrals.  It is important to remember that the modulation-strength variance $\sigma_A^2$ depends on the assumed model parameters $\boldsymbol\theta$, which set the values of the $A_n$. 

A hypothesis about the signal model can be compared against the background-only expectation by constructing a ratio of likelihood functions.  Take, for example, a Gaussian likelihood function $\mathcal{L}_{g}(x|N_i)$ with a single parameter $x$ and a  variance $\sigma^2(N_i)$, calculated from a sufficiently large data set $N_i$.  We might be interested in testing the null hypothesis given by $x=0$ against the alternative hypothesis with some arbitrary value for the parameter $x$.  In this case, the likelihood-ratio test statistic,
\begin{equation} \label{lfratio}
q_0(N_i) = - 2 \ln \left[\frac{\mathcal{L}_{g}\left(x=0 | N_i\right)}{\mathcal{L}_{g}\left(\hat{x}|N_i\right)}\right] \, , 
\end{equation}
is typically distributed according to a non-central chi-square probability distribution with one degree of freedom (DOF) over random realizations of the data set $N_i$~\cite{wald1943tests}.  Here, $\hat{x}$ is the parameter value that maximizes the likelihood for a given $N_i$.  That is to say, the test statistic $q_0$ is itself a random variable with a non-central single-DOF chi-square distribution; realizations of the data set $N_i$ that yield larger and less probable values of $q_0$ more strongly exclude the $x=0$ null hypothesis and thus support a claim of discovery.

The Asimov data set $\lambda_{ij,A}$ is one possibility in all realizations of $N_i$.  Therefore, we can apply the general result~\eqref{lfratio} to the specific Gaussian likelihood function $\mathcal{L}_{A,g}$($\xi$) in~\eqref{lfxigauss} with variance $\sigma_A$ as  in~\eqref{sigmaxi}.  This will allow us to test whether the typical Asimov data set $\lambda_{ij,A}$ excludes the $\xi=0$ null hypothesis of an unmodulated event rate at a desired significance level $\alpha$.  Specifically, this requires that ${q_0(\lambda_{ij,A}) = \sigma_A^{-2}}$ exceeds the single-DOF chi-square value corresponding to $\alpha$.  For example, ruling out the null hypothesis (no modulation) at a $95\%$ confidence level (corresponding to a single-DOF chi-square value of 3.84) and claiming a detection of the first-order mode requires an exposure of
\begin{equation}
M T \gtrsim 2 \times 3.84 \left[\int_{E_\text{thresh}}^{E_\text{max}}\! dE_\text{nr}\, \frac{A_1^2(E_\text{nr})}{A_0(E_\text{nr})}\right]^{-1} \equiv E(A_1)\,,
\end{equation}
for a given set of parameters $\boldsymbol\theta$.  We emphasize here that this is the exposure required to simply detect a modulation for specific DM model and experimental parameters.  Were we to allow the parameters $\boldsymbol\theta$ to be free, it would take a larger amount of exposure to constrain these parameters and characterize the model.  Studying the statistics of this characterization procedure in more detail is important in understanding the ability of the higher-harmonic modes to constrain the DM particle and astrophysical properties and is left to future work.

To derive the corresponding inequality for the $n^\text{th}$-order modes, the only changes required in the derivation are to let $A_1 \to A_n$ and $\cos\omega(t-t_0) \to \cos n\omega(t-t_0)$.  We see that the replacement of the discrete sum over time bins with an integral over $\cos^2 n\omega(t-t_0)$ in \eqref{sigmaxi} still yields a factor of $T/2$, so there are no further numerical factors introduced by the latter change.

Also of interest is the quantity $E(A_0)$, the minimum amount of exposure typically needed to detect the zeroth-order direct-detection signal at a background-free experiment.  In this work, we simply assume that observation of a single event constitutes detection, leading to the definition of $E(A_0)$ given in \eqref{EAEBapprox}.  A more careful analysis would take into consideration the expected background rates at each experiment and would result in straightforward modifications of the definitions of $E(A_0)$, $E(A_n)$, and $E(B_n)$.

\section{The Earth's Trajectory} \label{sec: ORBIT}

\subsection{Relative to the Sun}

In this Appendix, we sketch the derivation of the leading correction to the Earth's velocity coming from the eccentricity of the orbit.  We begin by considering the Earth's orbit in the ecliptic plane (see Fig.~\ref{EarthSun}), and 
we describe the orbit using the geometric construction of Kepler's laws.  The Earth follows a counterclockwise orbit around the Sun, constrained to the ellipse in Fig.~\ref{EarthSun}, with $a \approx 1.4960 \times 10^8$ km -- the length of the semi-major axis -- labeling the perihelion.  The Sun is located at one of the focal points, which is a distance $f = a \, e$ from the center of the ellipse, with $e \approx 0.016722$ the eccentricity of the orbit.  The eccentric anomaly is labeled $E$, and it is found geometrically using the procedure shown in Fig.~\ref{EarthSun}; it is the angle between the perihelion, the center of the ellipse, and a point on a fictitious circle of radius $a$.  If $t_p$ denotes the time of perihelion, then the angle $E$ evolves with time through the well-known relation
\es{meanAnom}{
g(t)= E - e \, \sin E \,, \qquad g(t) = \omega (t - t_p) \,.
}
$g(t)$ is commonly referred to as the mean anomaly. 
  The angle $\nu$, which is the angle between the perihelion, the Sun, and the Earth, is the true anomaly.  It is straightforward to show that through order $e^2$ in a small-$e$ expansion,
  \es{nuE}{
  \nu \approx g(t) + 2 \, e \sin g(t) + {5 \over 4} e^2 \sin 2 g(t) \,.
  }
  The distance from the Sun to the Earth at a given $\nu$ is simply 
  \es{SunR}{
  r(t) = {a (1 - e^2) \over 1+ e \cos \nu} \,.  
}

The ecliptic coordinate system is defined by the orthonormal unit vectors ${\V{ \hat \epsilon_1}}$ and ${\V{ \hat \epsilon_2}}$, which span the ecliptic plane and are simply related to the vernal equinox.\footnote{In 2013, the unit vectors are given in Galactic coordinates by 
\es{e12def}{
\V{\hat \epsilon_1} \approx (0.9940,0.1095,0.003116) \,, \qquad \V{\hat \epsilon_2} \approx (-0.05173, 0.4945, -0.8677) \,.
} 
}
At the vernal equinox, which occurs at time $t_1$, the vector ${\V{ \hat \epsilon_2}}$ points from the Earth to the Sun.  Note, for example, that in Fig.~\ref{EarthSun} the Earth has just passed the vernal equinox.  The projection of the Earth's rotational axis to the ecliptic plane is antiparallel to ${\V{ \hat \epsilon_1}}$.     

The ecliptic longitude $\lambda(t)$ of the Earth in the standard heliocentric ecliptic coordinates is measured counterclockwise with respect to the $ {\V{ \hat \epsilon_2}}$ axis; that is, $\lambda = 0$ at the autumnal equinox, and it evolves positively throughout the year ($\lambda = 180^\circ$ at the vernal equinox, $t = t_1$).  The ecliptic longitude is simply related to the true anomaly through the relation 
\es{lambdat}{
\lambda(t) = \lambda_p + \nu \,,
}
where 
$\lambda_p \approx 102^\circ$
 is the ecliptic longitude of the perihelion (in 2013).  The trajectory of the Earth throughout the year is then given by   
\es{trajectory}{
\V{r}(t) =r(t) \big( - \sin \lambda(t) \, \V{\hat \epsilon_1} + \cos \lambda(t) \, \V{\hat \epsilon_2} \big) \,.
}    

The velocity $\V{V_\earth}(t)$ is computed by evaluating $\dot{\V{r}}(t)$, and this may be done perturbatively in the small parameter $e$.
To zeroth order in $e$, we find
\es{vEcirc}{
\V{V}_\earth(t) \approx V_\earth \big( \V{\hat \epsilon_1} \cos \omega (t - t_1) + \V{\hat \epsilon_2} \sin \omega (t - t_1) \big) \,,
}
with $V_\earth \approx \omega \, a\, (1- e^2 / 4) \approx 29.79 \kms$ the mean speed of the Earth. 
At the next order in $e$ we get the expansion
\es{vEimproved}{
\V{V_\earth}(t) &\approx V_\earth \left[ \cos \omega (t - t_1) \big( \V{\hat \epsilon_1} - 2 \, e\, \sin \lambda_p \,  \V{\hat \epsilon_2}  \big) +
\sin \omega (t - t_1) \,\big( \V{\hat \epsilon_2} + 2 \, e\, \sin \lambda_p \,  \V{\hat \epsilon_1}  \big) \right. \\
&\left. -\, e \, \big( \cos 2 \omega (t - t_1)\, ( \cos \lambda_p \,\V{\hat \epsilon_1} -\sin \lambda_p\, \V{\hat \epsilon_2}) + \sin 2 \omega (t - t_1) \,(\sin \lambda_p \,\V{\hat \epsilon_1} +\cos \lambda_p\, \V{\hat \epsilon_2})\big) \right] \,.  
}

It is worth mentioning that the expansion~\eqref{vEimproved} differs in a key way from that given in~\cite{Lewin:1995rx}; the expansion in that work has propagated through many subsequent papers,
  and we believe that it is not completely correct.  

\subsection{Relative to the Galactic Center}

To leading order in $\epsilon$ and to zeroth order in the eccentricity $e$, we may add the velocity $\V{V_\earth}(t)$ to $\V{v_\Sun}$ and write 
\es{vobs1}{
v_\obs(t) = | \V{v_\Sun} + \V{V_\earth} (t)| \approx v_\Sun \left( 1 + \epsilon \, v_1 \, \cos \omega (t - t_0) \right) \,,
}
where $v_1$ is given in~\eqref{vbarvnGood} and $t_0 = t_1 + \phi$, with $\phi \approx 72.5$ days as in~\eqref{phin}.  At the next order in $\epsilon$, we must also include the leading corrections coming from the eccentricity, since $e / \epsilon \approx 0.52$.  This leads to the expansion in~\eqref{GoodExpand} -- \eqref{vbarvnGood}, and the time $t_0$ also receives a small shift: 
\es{t0approxF}{
t_0 \approx t_1 + \phi  + {4 \, e \over \omega} \cos\left(\lambda_p - {\omega \,\phi \over 2} \right) \sin {\omega \, \phi \over 2} \,,
}
which gives 
$t_0 \approx t_1 + 73.4$ days.
This is about 22 hours later than the time we found neglecting the eccentricity.  

It is also important to note that if we had instead used the incorrect expression for the Earth's velocity in~\cite{Lewin:1995rx}, then we would have found an extra factor of $3/2$ in the expressions for $v_2$, $u_1$, and $u_2$ in~\eqref{vbarvnGood}.

\subsection{Relative to the Earth's Center} \label{sec: ROT}

Figure~\ref{EarthSun}
illustrates how the unit vectors $\V{\hat \epsilon_1}$, $\V{\hat \epsilon_2}$, and $\V{\hat \epsilon_3} \equiv \V{\hat \epsilon}_1 \times \V{\hat \epsilon}_2$ are related to the ecliptic plane and the Earth's rotational axis.  The Earth's rotational axis is in the plane spanned by the vectors $\V{\hat \epsilon_1}$ and $\V{\hat \epsilon_3}$.  The angle $\varepsilon = 23.4^\circ$ is the obliquity of the Earth's axis, and it is the angle between the rotational axis and $\V{\hat \epsilon_3}$.  

With these considerations in mind, it is straightforward to write down an expression for the time-varying velocity $\V{V}_{(\phi_0,\lambda_0)}(t) $ of a point on the surface -- specified by the coordinates $(\phi_0, \lambda_0)$ -- relative to the Earth's center in terms of the unit vectors $\V{\hat \epsilon_1}$,  $\V{\hat \epsilon_2}$, and  $\V{\hat \epsilon_3}$;\footnote{The rotational speed $V_\text{d} \approx 0.463 \kms $ is found by multiplying $\omega_\text{d}$ by the mean radius of the Earth (we approximate the Earth by a sphere).}
\es{vEsurf}{\V{V}_{(\phi_0,\lambda_0)}(t) =
 - V_\text{d} \cos \phi_0 \big[ \sin \lambda_d(t) \, \V{\hat \epsilon_2} + \, \cos \lambda_d(t) \big( \cos \varepsilon \, \V{\hat \epsilon_1} + \sin \varepsilon \, \V{\hat \epsilon_3} \big)  \big] \,,
}
with $\lambda_d(t) = \omega_\text{d}\, (t - t_1) + \lambda_0$.
To accurately describe the trajectory of the observer on the surface of the Earth, we should add this velocity to $\V{v}_\Sun + \V{V}_\earth(t)$.  However, because we are studying a mode of frequency $\omega_d \gg \omega$, we may neglect $\V{V}_\earth(t)$ as a first approximation.  A straightforward calculation then leads to the result in~\eqref{dailyvobs}, with  
\es{udvd}{
t_0^d = t_1 + {1 \over \omega_d} \arccos \left({ \V{\hat v_\Sun} \cdot \V{ \hat \epsilon_1} \, \cos \varepsilon + \V{\hat v_\Sun} \cdot \V{ \hat \epsilon_3} \, \sin \varepsilon  \over d_d} \right) \approx t_1 +  2.9 \, \text{hours}.
}

  \section{Example Velocity Distributions} \label{Examples} \label{MaxSection}  \label{SHM}
  
  In this section, we calculate the first few Fourier coefficients explicitly as functions of $v_\Min$ -- using equations~\eqref{anbnGEN} -- for a few example isotropic velocity distributions.
  \subsection*{Maxwell Distribution} 

One of the most basic velocity distributions is the Maxwell distribution
\es{fmaxwell}{
\tilde f(\V{v}^2) = \left({1 \over \pi v_0^2 }\right)^{3/2} e^{- \V{v}^2 / v_0^2} \,.
}
The root-mean-squared DM speed is $\sigma_\text{RMS} = \sqrt{3/2} \, v_0$.  The mean inverse speed may be integrated exactly, giving
\es{etamaxwell}{
\eta(v_\Min,t) &= \left( { 1 \over \pi v_0^2} \right)^{3/2} 2 \pi \int_{-1}^1 ds \int_{v_\Min}^\infty dv \, v \, e^{- (v^2 + 2 \, s\, v \, v_\obs(t)  + v_\obs^2) / v_0^2 } \\
&=  {\erf(x+y) - \erf(x-y)\over 2 v_\obs(t)} \,,
}
where $v_\Min = x v_0$ and $v_\obs(t) = y v_0$.

To calculate the $a_n$ and $b_n$ coefficients, we may either substitute the expression for $v_\obs(t)$ in~\eqref{GoodExpand} into~\eqref{etamaxwell} and expand in $\epsilon$, or we may use the general formulas in~\eqref{anbnGEN}.  Either way, we find that 
\es{aMax}{
a_0 &= {\erf(x + \bar y) - \erf(x - \bar y) \over 2 \bar v }+ O(\epsilon^2) \,, \\
a_1 &=  v_1 \epsilon \left[ {1 \over \sqrt{\pi} \, v_0} \left( e^{-(x+\bar y)^2} + e^{-(x-\bar y)^2} \right) - a_0 \right]  +O(\epsilon^3) \,, \\
a_2 &= {\epsilon^2 \over 2} \left[ ( v_1^2 - 2  v_2) \left( a_0 - {1 \over \sqrt{\pi} \, v_0 } e^{-(x+\bar y)^2} \left(1 + e^{4 x \,\bar y } \right) \right) \right. \\
&\left.- { v_1^2\, \bar v \over \sqrt{\pi} \, v_0^2 } e^{-(x + \bar y)^2} \left( (\bar y + x) + e^{4 x \, \bar y} (\bar y - x) \right) \right] + O(\epsilon^4) \,,
}
where $\bar y \equiv \bar v / v_0$ and $b_1$, $b_2$ may be inferred from~\eqref{abfractions}. 
 
\subsection*{Standard Halo Model}

The SHM is similar to the Maxwell distribution, except $\tilde f(\V{v}) = 0$ when $|\V{v}| > v_\esc$, where $v_\esc$ is the escape velocity.  This attempts to take into account the fact that DM with sufficiently high velocity escapes the Galaxy.  

More explicitly, the SHM velocity distribution is given by 
\es{SHMf}{
\tilde f (\V{v}) = \left\{ \begin{array}{ll}
{1 \over N_\esc } \left( {1 \over \pi v_0^2 } \right)^{3/2} e^{- \V{v}^2 / v_0^2 } \,, \qquad &|\V{v}| < v_\esc \\
0 \,, \qquad &\text{else} \,,
\end{array}
\right.
}
with 
\es{Nesc}{
N_\esc = \erf(z) - {2 \over \sqrt{\pi}} z e^{-z^2} \,, \qquad z \equiv {v_\esc \over v_0} \,.
}
The mean inverse speed in the physical region $v_\esc > v_\text{obs}(t)$ is found to be
\es{SHMeta}{
\eta(v_\Min) = \left\{ \begin{array}{ll} 
( 2 N_\esc \, y \, v_0)^{-1} \left[ \erf(x+y) - \erf(x-y) - {4 \over \sqrt{\pi} } y e^{-z^2} \right] \,, & v_\Min < v_\esc - v_\text{obs}(t)  \\
( 2 N_\esc \, y \, v_0)^{-1}  \left[ \erf(z) - \erf(x-y) - {2 (y + z - x) \over \sqrt{\pi} } e^{-z^2} \right] \,, & \begin{array}{c} v_\esc - v_\obs(t) < v_\Min \\ < v_\esc + v_\obs(t) \end{array} \,. \\
0 & v_\Min > v_\esc + v_\obs(t) 
\end{array} \right.
}

The harmonic expansion in the SHM is complicated by the fact that the mean inverse speed is a piecewise function with the regions themselves depending on time.  We define the time-independent regions
\es{IandII}{
\text{(I)} \equiv v_\Min < v_\esc - \bar v \,, \qquad \text{(II)} \equiv v_\esc - \bar v < v_\Min < v_\esc + \bar v \,, \qquad \text{(III)} = v_\esc + \bar v < v_\Min \,,
}
and in each of these regions we expand the $a_0$, $a_1$, $a_2$, $b_1$, and $b_2$ harmonic coefficients to leading order in $\epsilon$:
\es{aSHO}{
a_0 &\approx \left\{  \begin{array}{lr}
{1 \over 2 \bar v N_\esc(z)} \left( \erf(x+\bar y) - \erf(x-\bar y) - {4 \bar y \over \sqrt{\pi}} e^{-z^2} \right)   \, & \quad\quad\quad\quad\quad\quad\quad\quad\quad\quad\quad \quad\quad \text{(I)}  \\ 
 {1 \over 2 \bar v N_\esc(z)} \left(\erf(z) - \erf(x-\bar y) - {2 (z + \bar y - x) \over \sqrt{\pi} } e^{-z^2} \right)    \, & \quad\quad\quad\quad\quad\quad\quad\quad\quad\quad\quad \quad\quad\text{(II)}  \\
\begin{array}{l} 0 \end{array} \, & \quad\quad\quad\quad\quad \quad\quad\text{(III)}
\end{array} \,    \right. \\
a_1 &\approx \left\{ \begin{array}{lr}
{ v_1 \epsilon \over \bar v N_\esc(z)}  \left[ {\bar y \over \sqrt{\pi}} \left( e^{-(x+\bar y)^2} + e^{-(x-\bar y)^2} \right) - \left({ \erf(x+\bar y) - \erf(x-\bar y) \over 2}\right) \right]  \, & \quad\quad\quad\quad \quad\quad\,\text{(I)}\\ 
{ v_1 \epsilon \over 2 \bar v N_\esc(z)} \left[ \erf(x-\bar y) -\erf(z) +  {2 \over \sqrt{\pi}} \left( \bar y e^{-(x-\bar y)^2} -  (x- z) e^{-z^2} \right) \right]  \,& \text{(II)}\\
\begin{array}{l} 0 \end{array} \,& \text{(III)}\\
\end{array} \right.  \\
a_2 &\approx \left\{~\begin{array}{lr}
\begin{array}{lr}
 {\epsilon^2 \over 2 \bar v N_\esc(z)} \left[ ( v_1^2 - 2  v_2) \left( {1 \over 2 } \left( \erf(x+\bar y) - \erf(x-\bar y) \right) \right. \right. \\
 \left. \left. - {\bar y \over \sqrt{\pi}  } e^{-(x+\bar y)^2} \left(1 + e^{4 x \cdot \bar y } \right) \right) - { v_1^2 \bar y^2 \over \sqrt{\pi} } e^{-(x + \bar y)^2} \left( (x+\bar y) - (x-\bar y) e^{4 x \cdot \bar y } \right) \right]  
\end{array} \,& \quad\quad\quad\text{(I)}\\
\begin{array}{lr}
{\epsilon^2 \over 2 \bar v N_\esc(z)} \left[ ( v_1^2 - 2  v_2) \left( {1 \over 2 } \left( \erf(z) - \erf(x-\bar y) \right) \right. \right. \\
 \left. \left. - {1 \over \sqrt{\pi}  } \left(\bar y e^{-(x-\bar y)^2} + (z - x) e^{- z^2 } \right) \right) + { v_1^2 \bar y^2 \over \sqrt{\pi} } e^{-(x - \bar y)^2} \left( x - \bar y \right) \right] 
\end{array}\,& \text{(II)}\\
\begin{array}{l}0 \end{array} \,& \quad\quad\,\,\, \text{(III)}
\end{array} \right.  \\ 
b_1 &\approx { u_1 \over v_1} \epsilon\, a_1  \,, \qquad b_2 \approx  { u_2 \over  v_1} \epsilon \, a_1  \,.
} 
The SHM is well approximated by the Maxwell distribution for velocities less than $v_\esc$.

\subsection*{Dark-Matter Stream} 

For a velocity distribution given by a delta function, the mean inverse speed is
  \es{etaStream}{
\eta(v_\Min,t) = {1 \over v_\obs^S(t)} \theta\big( v_\obs^S(t) - v_\Min \big) \,.
}
  It is convenient to consider $\eta(v_\Min,t)$ in three different regions, defined by 
\es{StreamRegions}{
(\text{I}) &= \{v_\Min < \text{Min} \left[ v_\obs^S(t) \right] \} \,, \qquad (\text{II}) = \{\text{Min} \left[ v_\obs^S(t) \right]  < v_\Min < \text{Max} \left[ v_\obs^S(t) \right]  \} \\
(\text{III}) &= \{ \text{Max} \left[ v_\obs^S(t) \right]  < v_\Min  \} \,,
}
where $\text{Min} \left[ v_\obs^S(t) \right]$ and $\text{Max} \left[ v_\obs^S(t) \right]$ denote the minimum and maximum values, respectively, of $v_\obs^S(t)$ over a complete period.  In (I), the mean inverse speed is simply given by the smooth and continuous function $1 /v_\obs^S(t)$, and the methods described in Sec.~\ref{sec: mis} apply.  In (III), the mean inverse speed vanishes identically, and so the only new, nontrivial dynamics occurs within (II).  In this region, the mean inverse speed generically has two discontinuities during a full year -- that is, it looks like a slightly modulated box function (see Fig.~\ref{AnMod}).  
To leading order in $\epsilon_S$ we may approximate region (II) by 
\es{regionIIapprox}{
\text{(II)} \approx \{ v_\Min : \bar v^S (1 - \epsilon_S |v_1^S|) < v_\Min < \bar v^S(1 + \epsilon_S | v_1^S |) \} \,.
}
In the example of a stream traveling at a speed of $350 \kms$ towards the south Galactic pole this corresponds to the narrow range $v_\Min \in (412, 440) \kms$.  

With $v_\Min$ in region (II), it is relatively straightforward to approximate the $a_n^S$ and $b_n^S$ harmonic coefficients to leading order in $\epsilon_S$.  First, let $t_a$ and $t_b$, with $t_a < t_b$, be the two times during the year when $v_\obs^S(t) = v_\Min$.  These times (relative to $t_0^S$) may be approximated by 
\es{t1t2}{
t_a &=  {1 \over \omega} \left[  \cos^{-1} \left[ {v_\Min - \bar v^S \over \epsilon_S \, v_1^S \, \bar v^S} \right] + O(\epsilon_S) \right] \,, \\
t_b &=  {1 \over \omega} \left[2 \pi -  \cos^{-1} \left[ {v_\Min - \bar v^S \over \epsilon_S \, v_1^S \, \bar v^S} \right] + O(\epsilon_S) \right] \,.
}
We then approximate $v_\obs^S(t) \approx \bar v^S$ in the denominator of~\eqref{etaStream} and find
\es{regionIIStream}{
a_0^S(v_\Min) &\approx {1 \over \bar v^S} \left[ \theta(v_1^S) - {\omega \over 2 \pi} \text{sign}(v_1^S) (t_b - t_a) \right] \,,\\
a_n^S(v_\Min) &\approx -{1 \over \bar v^S} {\sin (n \, \omega \,t_b) - \sin (n \, \omega \, t_a) \over n \pi} \text{sign} (v_1^S)\,,  \\
b_n^S(v_\Min) &\approx {1 \over \bar v^S}  {\cos (n \, \omega \,t_b) - \cos (n \, \omega \, t_a) \over n \pi} \text{sign} (v_1^S)\,.  
}  
Using~\eqref{t1t2}, we then see that the $a_n$ are order $\epsilon_S^0$ and given by the expression in~\eqref{anFinStream} while the $b_n$ are suppressed by $O(\epsilon_S)$.

  \clearpage

\bibliographystyle{ssg}
\bibliography{DMmod}

\begingroup\raggedright\begin{thebibliography}{10}

\bibitem{Komatsu:2010fb}
{\bf WMAP Collaboration} Collaboration, E.~Komatsu {\em et.~al.}, ``{Seven-Year
  Wilkinson Microwave Anisotropy Probe (WMAP) Observations: Cosmological
  Interpretation},'' {\em Astrophys.J.Suppl.} {\bf 192} (2011) 18,
  \href{http://xxx.lanl.gov/abs/1001.4538}{{\tt 1001.4538}}.

\bibitem{Ade:2013zuv}
{\bf Planck Collaboration} Collaboration, P.~Ade {\em et.~al.}, ``{Planck 2013
  results. XVI. Cosmological parameters},''
  \href{http://xxx.lanl.gov/abs/1303.5076}{{\tt 1303.5076}}.

\bibitem{Goodman:1984dc}
M.~W. Goodman and E.~Witten, ``{Detectability of Certain Dark Matter
  Candidates},'' {\em Phys.Rev.} {\bf D31} (1985) 3059.

\bibitem{Drukier:1983gj}
A.~Drukier and L.~Stodolsky, ``{Principles and Applications of a Neutral
  Current Detector for Neutrino Physics and Astronomy},'' {\em Phys.Rev.} {\bf
  D30} (1984) 2295.

\bibitem{Smith:1988kw}
P.~Smith and J.~Lewin, ``{Dark Matter Detection},'' {\em Phys.Rept.} {\bf 187}
  (1990) 203.

\bibitem{Bertone:2004pz}
G.~Bertone, D.~Hooper, and J.~Silk, ``{Particle dark matter: Evidence,
  candidates and constraints},'' {\em Phys.Rept.} {\bf 405} (2005) 279--390,
  \href{http://xxx.lanl.gov/abs/hep-ph/0404175}{{\tt hep-ph/0404175}}.

\bibitem{Jungman:1995df}
G.~Jungman, M.~Kamionkowski, and K.~Griest, ``{Supersymmetric dark matter},''
  {\em Phys.Rept.} {\bf 267} (1996) 195--373,
  \href{http://xxx.lanl.gov/abs/hep-ph/9506380}{{\tt hep-ph/9506380}}.

\bibitem{Gaitskell:2004gd}
R.~Gaitskell, ``{Direct detection of dark matter},'' {\em
  Ann.Rev.Nucl.Part.Sci.} {\bf 54} (2004) 315--359.

\bibitem{Freese:2012xd}
K.~Freese, M.~Lisanti, and C.~Savage, ``{Annual Modulation of Dark Matter: A
  Review},'' \href{http://xxx.lanl.gov/abs/1209.3339}{{\tt 1209.3339}}.

\bibitem{Drukier:1986tm}
A.~Drukier, K.~Freese, and D.~Spergel, ``{Detecting Cold Dark Matter
  Candidates},'' {\em Phys.Rev.} {\bf D33} (1986) 3495--3508.

\bibitem{Gelmini:2000dm}
G.~Gelmini and P.~Gondolo, ``{WIMP annual modulation with opposite phase in
  Late-Infall halo models},'' {\em Phys.Rev.} {\bf D64} (2001) 023504,
  \href{http://xxx.lanl.gov/abs/hep-ph/0012315}{{\tt hep-ph/0012315}}.

\bibitem{Freese:2003na}
K.~Freese, P.~Gondolo, H.~J. Newberg, and M.~Lewis, ``{The effects of the
  Sagittarius dwarf tidal stream on dark matter detectors},'' {\em
  Phys.Rev.Lett.} {\bf 92} (2004) 111301,
  \href{http://xxx.lanl.gov/abs/astro-ph/0310334}{{\tt astro-ph/0310334}}.

\bibitem{Savage:2006qr}
C.~Savage, K.~Freese, and P.~Gondolo, ``{Annual Modulation of Dark Matter in
  the Presence of Streams},'' {\em Phys.Rev.} {\bf D74} (2006) 043531,
  \href{http://xxx.lanl.gov/abs/astro-ph/0607121}{{\tt astro-ph/0607121}}.

\bibitem{Fornengo:2003fm}
N.~Fornengo and S.~Scopel, ``{Temporal distortion of annual modulation at low
  recoil energies},'' {\em Phys.Lett.} {\bf B576} (2003) 189--194,
  \href{http://xxx.lanl.gov/abs/hep-ph/0301132}{{\tt hep-ph/0301132}}.

\bibitem{Green:2003yh}
A.~M. Green, ``{Effect of realistic astrophysical inputs on the phase and shape
  of the WIMP annual modulation signal},'' {\em Phys.Rev.} {\bf D68} (2003)
  023004, \href{http://xxx.lanl.gov/abs/astro-ph/0304446}{{\tt
  astro-ph/0304446}}.

\bibitem{Spergel:1987kx}
D.~N. Spergel, ``{THE MOTION OF THE EARTH AND THE DETECTION OF WIMPs},'' {\em
  Phys.Rev.} {\bf D37} (1988) 1353.

\bibitem{Alves:2010pt}
D.~S. Alves, M.~Lisanti, and J.~G. Wacker, ``{Poker face of inelastic dark
  matter: Prospects at upcoming direct detection experiments},'' {\em
  Phys.Rev.} {\bf D82} (2010) 031901,
  \href{http://xxx.lanl.gov/abs/1005.5421}{{\tt 1005.5421}}.

\bibitem{Chang:2011eb}
S.~Chang, J.~Pradler, and I.~Yavin, ``{Statistical Tests of Noise and Harmony
  in Dark Matter Modulation Signals},'' {\em Phys.Rev.} {\bf D85} (2012)
  063505, \href{http://xxx.lanl.gov/abs/1111.4222}{{\tt 1111.4222}}.

\bibitem{Lewin:1995rx}
J.~Lewin and P.~Smith, ``{Review of mathematics, numerical factors, and
  corrections for dark matter experiments based on elastic nuclear recoil},''
  {\em Astropart.Phys.} {\bf 6} (1996) 87--112.

\bibitem{Schoenrich:2009bx}
R.~Schoenrich, J.~Binney, and W.~Dehnen, ``{Local Kinematics and the Local
  Standard of Rest},'' \href{http://xxx.lanl.gov/abs/0912.3693}{{\tt
  0912.3693}}.

\bibitem{2000A&A...354..522M}
F.~Mignard, ``Local galactic kinematics from Hipparcos proper motions,'' {\em
  Astron. Astrophys.} {\bf 354} (Feb., 2000) 522--536.

\bibitem{Alfonsi:2012bfa}
{\bf XENON Collaboration} Collaboration, M.~Alfonsi, ``{The XENON Dark Matter
  programme: From XENON100 to XENON1T},'' {\em PoS} {\bf DSU2012} (2012) 047.

\bibitem{FigueroaFeliciano:2010zz}
{\bf CDMS Collaboration} Collaboration, E.~Figueroa-Feliciano, ``{Towards
  direct detection of WIMPs with the cryogenic dark matter search},'' {\em AIP
  Conf.Proc.} {\bf 1200} (2010) 959--962.

\bibitem{Akerib:2012ys}
{\bf LUX Collaboration} Collaboration, D.~Akerib {\em et.~al.}, ``{The Large
  Underground Xenon (LUX) Experiment},'' {\em Nucl.Instrum.Meth.} {\bf A704}
  (2013) 111--126, \href{http://xxx.lanl.gov/abs/1211.3788}{{\tt 1211.3788}}.

\bibitem{Aprile:2011dd}
{\bf XENON100 Collaboration} Collaboration, E.~Aprile {\em et.~al.}, ``{The
  XENON100 Dark Matter Experiment},'' {\em Astropart.Phys.} {\bf 35} (2012)
  573--590, \href{http://xxx.lanl.gov/abs/1107.2155}{{\tt 1107.2155}}.

\bibitem{Aprile:2012nq}
{\bf XENON100 Collaboration} Collaboration, E.~Aprile {\em et.~al.}, ``{Dark
  Matter Results from 225 Live Days of XENON100 Data},'' {\em Phys.Rev.Lett.}
  {\bf 109} (2012) 181301, \href{http://xxx.lanl.gov/abs/1207.5988}{{\tt
  1207.5988}}.

\bibitem{Agnese:2013rvf}
{\bf CDMS Collaboration} Collaboration, R.~Agnese {\em et.~al.}, ``{Dark Matter
  Search Results Using the Silicon Detectors of CDMS II},'' {\em
  Phys.Rev.Lett.} (2013) \href{http://xxx.lanl.gov/abs/1304.4279}{{\tt
  1304.4279}}.

\bibitem{Caldwell:1981rj}
J.~Caldwell and J.~Ostriker, ``{The Mass distribution within our Galaxy: A
  Three component model},'' {\em Astrophys.J.} {\bf 251} (1981) 61--87.

\bibitem{Catena:2009mf}
R.~Catena and P.~Ullio, ``{A novel determination of the local dark matter
  density},'' {\em JCAP} {\bf 1008} (2010) 004,
  \href{http://xxx.lanl.gov/abs/0907.0018}{{\tt 0907.0018}}.

\bibitem{Weber:2009pt}
M.~Weber and W.~de~Boer, ``{Determination of the Local Dark Matter Density in
  our Galaxy},'' {\em Astron.Astrophys.} {\bf 509} (2010) A25,
  \href{http://xxx.lanl.gov/abs/0910.4272}{{\tt 0910.4272}}.

\bibitem{Salucci:2010qr}
P.~Salucci, F.~Nesti, G.~Gentile, and C.~Martins, ``{The dark matter density at
  the Sun's location},'' {\em Astron.Astrophys.} {\bf 523} (2010) A83,
  \href{http://xxx.lanl.gov/abs/1003.3101}{{\tt 1003.3101}}.

\bibitem{Pato:2010yq}
M.~Pato, O.~Agertz, G.~Bertone, B.~Moore, and R.~Teyssier, ``{Systematic
  uncertainties in the determination of the local dark matter density},'' {\em
  Phys.Rev.} {\bf D82} (2010) 023531,
  \href{http://xxx.lanl.gov/abs/1006.1322}{{\tt 1006.1322}}.

\bibitem{Helm:1956zz}
R.~H. Helm, ``{Inelastic and Elastic Scattering of 187-Mev Electrons from
  Selected Even-Even Nuclei},'' {\em Phys.Rev.} {\bf 104} (1956) 1466--1475.

\bibitem{Lewis:2003bv}
M.~J. Lewis and K.~Freese, ``{The Phase of the annual modulation: Constraining
  the WIMP mass},'' {\em Phys.Rev.} {\bf D70} (2004) 043501,
  \href{http://xxx.lanl.gov/abs/astro-ph/0307190}{{\tt astro-ph/0307190}}.

\bibitem{Cherwinka:2011ij}
J.~Cherwinka, R.~Co, D.~Cowen, D.~Grant, F.~Halzen, {\em et.~al.}, ``{A Search
  for the Dark Matter Annual Modulation in South Pole Ice},'' {\em
  Astropart.Phys.} {\bf 35} (2012) 749--754,
  \href{http://xxx.lanl.gov/abs/1106.1156}{{\tt 1106.1156}}.

\bibitem{Read:2008fh}
J.~Read, G.~Lake, O.~Agertz, and V.~P. Debattista, ``{Thin, thick and dark
  discs in LCDM},'' \href{http://xxx.lanl.gov/abs/0803.2714}{{\tt 0803.2714}}.

\bibitem{Read:2009iv}
J.~Read, L.~Mayer, A.~Brooks, F.~Governato, and G.~Lake, ``{A dark matter disc
  in three cosmological simulations of Milky Way mass galaxies},''
  \href{http://xxx.lanl.gov/abs/0902.0009}{{\tt 0902.0009}}.

\bibitem{Purcell:2009yp}
C.~W. Purcell, J.~S. Bullock, and M.~Kaplinghat, ``{The Dark Disk of the Milky
  Way},'' {\em Astrophys.J.} {\bf 703} (2009) 2275--2284,
  \href{http://xxx.lanl.gov/abs/0906.5348}{{\tt 0906.5348}}.

\bibitem{Bruch:2008rx}
T.~Bruch, J.~Read, L.~Baudis, and G.~Lake, ``{Detecting the Milky Way's Dark
  Disk},'' {\em Astrophys.J.} {\bf 696} (2009) 920--923,
  \href{http://xxx.lanl.gov/abs/0804.2896}{{\tt 0804.2896}}.

\bibitem{Diemand:2006ik}
J.~Diemand, M.~Kuhlen, and P.~Madau, ``{Dark matter substructure and gamma-ray
  annihilation in the Milky Way halo},'' {\em Astrophys.J.} {\bf 657} (2007)
  262--270, \href{http://xxx.lanl.gov/abs/astro-ph/0611370}{{\tt
  astro-ph/0611370}}.

\bibitem{Freese:2003tt}
K.~Freese, P.~Gondolo, and H.~J. Newberg, ``{Detectability of weakly
  interacting massive particles in the Sagittarius dwarf tidal stream},'' {\em
  Phys.Rev.} {\bf D71} (2005) 043516,
  \href{http://xxx.lanl.gov/abs/astro-ph/0309279}{{\tt astro-ph/0309279}}.

\bibitem{Aprile:2010bt}
{\bf XENON Collaboration} Collaboration, E.~Aprile {\em et.~al.}, ``{Design and
  Performance of the XENON10 Dark Matter Experiment},'' {\em Astropart.Phys.}
  {\bf 34} (2011) 679--698, \href{http://xxx.lanl.gov/abs/1001.2834}{{\tt
  1001.2834}}.

\bibitem{Angle:2007uj}
{\bf XENON Collaboration} Collaboration, J.~Angle {\em et.~al.}, ``{First
  Results from the XENON10 Dark Matter Experiment at the Gran Sasso National
  Laboratory},'' {\em Phys.Rev.Lett.} {\bf 100} (2008) 021303,
  \href{http://xxx.lanl.gov/abs/0706.0039}{{\tt 0706.0039}}.

\bibitem{Angle:2011th}
{\bf XENON10 Collaboration} Collaboration, J.~Angle {\em et.~al.}, ``{A search
  for light dark matter in XENON10 data},'' {\em Phys.Rev.Lett.} {\bf 107}
  (2011) 051301, \href{http://xxx.lanl.gov/abs/1104.3088}{{\tt 1104.3088}}.

\bibitem{Akerib:2005zy}
{\bf CDMS Collaboration} Collaboration, D.~Akerib {\em et.~al.}, ``{Exclusion
  limits on the WIMP-nucleon cross section from the first run of the Cryogenic
  Dark Matter Search in the Soudan Underground Laboratory},'' {\em Phys.Rev.}
  {\bf D72} (2005) 052009, \href{http://xxx.lanl.gov/abs/astro-ph/0507190}{{\tt
  astro-ph/0507190}}.

\bibitem{Ahmed:2009zw}
{\bf CDMS-II Collaboration} Collaboration, Z.~Ahmed {\em et.~al.}, ``{Dark
  Matter Search Results from the CDMS II Experiment},'' {\em Science} {\bf 327}
  (2010) 1619--1621, \href{http://xxx.lanl.gov/abs/0912.3592}{{\tt 0912.3592}}.

\bibitem{Ahmed:2010wy}
{\bf CDMS-II Collaboration} Collaboration, Z.~Ahmed {\em et.~al.}, ``{Results
  from a Low-Energy Analysis of the CDMS II Germanium Data},'' {\em
  Phys.Rev.Lett.} {\bf 106} (2011) 131302,
  \href{http://xxx.lanl.gov/abs/1011.2482}{{\tt 1011.2482}}.

\bibitem{Ahmed:2012vq}
{\bf CDMS Collaboration} Collaboration, Z.~Ahmed {\em et.~al.}, ``{Search for
  annual modulation in low-energy CDMS-II data},''
  \href{http://xxx.lanl.gov/abs/1203.1309}{{\tt 1203.1309}}.

\bibitem{Ahmed:2011gh}
{\bf CDMS Collaboration, EDELWEISS Collaboration} Collaboration, Z.~Ahmed {\em
  et.~al.}, ``{Combined Limits on WIMPs from the CDMS and EDELWEISS
  Experiments},'' {\em Phys.Rev.} {\bf D84} (2011) 011102,
  \href{http://xxx.lanl.gov/abs/1105.3377}{{\tt 1105.3377}}.

\bibitem{Armengaud:2012pfa}
{\bf EDELWEISS Collaboration} Collaboration, E.~Armengaud {\em et.~al.}, ``{A
  search for low-mass WIMPs with EDELWEISS-II heat-and-ionization detectors},''
  {\em Phys.Rev.} {\bf D86} (2012) 051701,
  \href{http://xxx.lanl.gov/abs/1207.1815}{{\tt 1207.1815}}.

\bibitem{Bernabei:2008yh}
{\bf DAMA Collaboration} Collaboration, R.~Bernabei {\em et.~al.}, ``{The
  DAMA/LIBRA apparatus},'' {\em Nucl.Instrum.Meth.} {\bf A592} (2008) 297--315,
  \href{http://xxx.lanl.gov/abs/0804.2738}{{\tt 0804.2738}}.

\bibitem{Bernabei:2010mq}
{\bf DAMA Collaboration, LIBRA Collaboration} Collaboration, R.~Bernabei {\em
  et.~al.}, ``{New results from DAMA/LIBRA},'' {\em Eur.Phys.J.} {\bf C67}
  (2010) 39--49, \href{http://xxx.lanl.gov/abs/1002.1028}{{\tt 1002.1028}}.

\bibitem{Bottino:2003iu}
A.~Bottino, F.~Donato, N.~Fornengo, and S.~Scopel, ``{Lower bound on the
  neutralino mass from new data on CMB and implications for relic
  neutralinos},'' {\em Phys.Rev.} {\bf D68} (2003) 043506,
  \href{http://xxx.lanl.gov/abs/hep-ph/0304080}{{\tt hep-ph/0304080}}.

\bibitem{Bottino:2003cz}
A.~Bottino, F.~Donato, N.~Fornengo, and S.~Scopel, ``{Light neutralinos and
  WIMP direct searches},'' {\em Phys.Rev.} {\bf D69} (2004) 037302,
  \href{http://xxx.lanl.gov/abs/hep-ph/0307303}{{\tt hep-ph/0307303}}.

\bibitem{Gondolo:2005hh}
P.~Gondolo and G.~Gelmini, ``{Compatibility of DAMA dark matter detection with
  other searches},'' {\em Phys.Rev.} {\bf D71} (2005) 123520,
  \href{http://xxx.lanl.gov/abs/hep-ph/0504010}{{\tt hep-ph/0504010}}.

\bibitem{Petriello:2008jj}
F.~Petriello and K.~M. Zurek, ``{DAMA and WIMP dark matter},'' {\em JHEP} {\bf
  0809} (2008) 047, \href{http://xxx.lanl.gov/abs/0806.3989}{{\tt 0806.3989}}.

\bibitem{Chang:2008xa}
S.~Chang, A.~Pierce, and N.~Weiner, ``{Using the Energy Spectrum at DAMA/LIBRA
  to Probe Light Dark Matter},'' {\em Phys.Rev.} {\bf D79} (2009) 115011,
  \href{http://xxx.lanl.gov/abs/0808.0196}{{\tt 0808.0196}}.

\bibitem{Savage:2008er}
C.~Savage, G.~Gelmini, P.~Gondolo, and K.~Freese, ``{Compatibility of
  DAMA/LIBRA dark matter detection with other searches},'' {\em JCAP} {\bf
  0904} (2009) 010, \href{http://xxx.lanl.gov/abs/0808.3607}{{\tt 0808.3607}}.

\bibitem{Aalseth:2010vx}
{\bf CoGeNT collaboration} Collaboration, C.~Aalseth {\em et.~al.}, ``{Results
  from a Search for Light-Mass Dark Matter with a P-type Point Contact
  Germanium Detector},'' {\em Phys.Rev.Lett.} {\bf 106} (2011) 131301,
  \href{http://xxx.lanl.gov/abs/1002.4703}{{\tt 1002.4703}}.

\bibitem{Aalseth:2011wp}
C.~Aalseth, P.~Barbeau, J.~Colaresi, J.~Collar, J.~Diaz~Leon, {\em et.~al.},
  ``{Search for an Annual Modulation in a P-type Point Contact Germanium Dark
  Matter Detector},'' {\em Phys.Rev.Lett.} {\bf 107} (2011) 141301,
  \href{http://xxx.lanl.gov/abs/1106.0650}{{\tt 1106.0650}}.

\bibitem{Aalseth:2012if}
{\bf CoGeNT Collaboration} Collaboration, C.~Aalseth {\em et.~al.}, ``{CoGeNT:
  A Search for Low-Mass Dark Matter using p-type Point Contact Germanium
  Detectors},'' \href{http://xxx.lanl.gov/abs/1208.5737}{{\tt 1208.5737}}.

\bibitem{Frandsen:2011ts}
M.~T. Frandsen, F.~Kahlhoefer, J.~March-Russell, C.~McCabe, M.~McCullough, {\em
  et.~al.}, ``{On the DAMA and CoGeNT Modulations},'' {\em Phys.Rev.} {\bf D84}
  (2011) 041301, \href{http://xxx.lanl.gov/abs/1105.3734}{{\tt 1105.3734}}.

\bibitem{Hooper:2011hd}
D.~Hooper and C.~Kelso, ``{Implications of CoGeNT's New Results For Dark
  Matter},'' {\em Phys.Rev.} {\bf D84} (2011) 083001,
  \href{http://xxx.lanl.gov/abs/1106.1066}{{\tt 1106.1066}}.

\bibitem{Fox:2011px}
P.~J. Fox, J.~Kopp, M.~Lisanti, and N.~Weiner, ``{A CoGeNT Modulation
  Analysis},'' {\em Phys.Rev.} {\bf D85} (2012) 036008,
  \href{http://xxx.lanl.gov/abs/1107.0717}{{\tt 1107.0717}}.

\bibitem{Lee:2013wza}
S.~K. Lee, M.~Lisanti, A.~H.~G. Peter, and B.~R. Safdi, ``{The Effect of
  Gravitational Focusing on Annual Modulation},''
  \href{http://xxx.lanl.gov/abs/1308.1953}{{\tt 1308.1953}}.

\bibitem{Alenazi:2006wu}
M.~S. Alenazi and P.~Gondolo, ``{Phase-space distribution of unbound dark
  matter near the Sun},'' {\em Phys.Rev.} {\bf D74} (2006) 083518,
  \href{http://xxx.lanl.gov/abs/astro-ph/0608390}{{\tt astro-ph/0608390}}.

\bibitem{Cowan:2010js}
G.~Cowan, K.~Cranmer, E.~Gross, and O.~Vitells, ``{Asymptotic formulae for
  likelihood-based tests of new physics},'' {\em Eur.Phys.J.} {\bf C71} (2011)
  1554, \href{http://xxx.lanl.gov/abs/1007.1727}{{\tt 1007.1727}}.

\bibitem{wald1943tests}
A.~Wald, ``Tests of statistical hypotheses concerning several parameters when
  the number of observations is large,'' {\em Transactions of the American
  Mathematical Society} {\bf 54} (1943), no.~3 426--482.

\end{thebibliography}\endgroup

\end{document}